\documentclass[aps,prd,preprintnumbers,superscriptaddress,nofootinbib]{revtex4}

\usepackage{hyperref}
\hypersetup{colorlinks=true,linkcolor=purple,anchorcolor=blue,citecolor=blue, filecolor=blue,urlcolor=red,bookmarksnumbered=true,
pdfview=FitB
}
\usepackage{color}
\usepackage{xcolor}
\colorlet{purple1}{blue!70!red}
\colorlet{darkred}{red!50!black}

\usepackage{graphicx}
\usepackage[bf,SL,BF]{subfigure}
\usepackage{psfrag}
\usepackage{color}
\usepackage{amssymb}
\usepackage{amsmath}
\usepackage{epstopdf}
\usepackage{bbold}
\usepackage[normalem]{ulem}

\newcommand{\be}{\begin{eqnarray}}
\newcommand{\ee}{\end{eqnarray}}

\newcommand{\bfb}{{\bf b}_{\perp}}

\newcommand{\bfp}{{\bf p}_{\perp}}  
\newcommand{\pp}{{p}_{\perp}}  
\newcommand{\dpi}{{\Delta}_{\perp}}  

\newcommand{\bfd}{{\bf \Delta}_{\perp}} 
\newcommand{\Dp}{{\bf \Delta}_{\perp}}

\newcommand{\bfD}{{\bf D}_{\perp}}

\newcommand{\Aodp}{A^\nu_1(x^{\prime\prime})}
\newcommand{\Aop}{A^\nu_1(x^{\prime})}
\newcommand{\Atdp}{A^\nu_2(x^{\prime\prime})}
\newcommand{\Atp}{A^\nu_2(x^{\prime})}
\newcommand{\zf}{\sqrt{1-\xi^2}}
\newcommand{\zfs}{(1-\xi^2)}
\newcommand{\exf}{\exp\big[-\tilde{a}(x^{\prime\prime})\bfp^{\prime\prime 2}-\tilde{a}(x^{\prime})\bfp^{\prime 2} \big]}

\begin{document}


\title{T-odd Wigner Distributions in boost-invariant longitudinal position space and Spin-momentum correlation in proton }


\author{Tanmay~Maji}
\email{tanmayphy@nitkkr.ac.in} 
\affiliation{Department of Physics, National Institute of Technology Kurukshetra, India 136119}

\date{\today}

\begin{abstract}
The boost invariant longitudinal position space variable $\sigma$, which is the Fourier conjugate to skewness $\xi$, unravels the longitudinal impact parameter information in a proton. Here, we investigate the skewness sensitivity of T-odd leading twist GTMDs within the Dokshitzer Gribov Lipatov Altarelli Parisi (DGLAP) region, considering a momentum transfer to longitudinal as well as transverse direction. 
The $\sigma$-space distributions of the T-odd sector show oscillatory patterns that are sensitive to the square of the total momentum transfer $-t$, which is analogous to the diffraction scattering of waves in Optics. An additional effect on the diffraction pattern is reported, caused by interference between transverse momentum transfer $\bfd$ to the transverse momentum $\bfp$ of quarks. We also present the correlation of proton spin to the transverse momentum of constituents through Sivers and Boer-Mulders Wigner Distributions in boost invariant longitudinal position space.

\end{abstract}

\maketitle

\section{Introduction\label{intro}}
The non-perturbative structure of hadrons can be understood by investigating parton distributions of the constituents, quarks and gluons. In recent times, the most generic Wigner distributions(WDs), introduced by Ji \cite{Ji:2003ak,Belitsky:2003nz}, and  Generalized Transverse Momentum Distributions(GTMDs), introduced in \cite{Meissner:2008ay,Meissner:2009ww,Echevarria:2016mrc}, have drawn attention, as these provide a generic tomography of hadrons. The Wigner distribution is a quasi-probabilistic six-dimensional distribution in phase space, providing a comprehensive parton structure of the proton by incorporating both spatial and momentum distributions. Such distributions are not measured directly in experiments. However, WDs reduce to Generalized Parton Distributions(GPDs) under integration over transverse momenta, and transverse impact parameter integration provides Transverse Momentum Distributions(TMDs). The phase space average of WDs gives rise to spin-spin and spin-orbital angular momentum (OAM) correlations between a nucleon and its constituents.  The GPDs appear in hard exclusive reactions such as deeply virtual Compton scattering DVCS) or deeply virtual meson production(DVMP), whereas the TMDs are accessible in semi-inclusive deep inelastic scattering(SIDIS) and Drell-Yan process \cite{Mulders:1995dh,Barone:2001sp,Bacchetta:2006tn,Brodsky:2002cx,Bacchetta:2017gcc}.
The T-even quark Wigner distributions at leading twist have been studied in different models e.g., in chiral soliton model\cite{Lorce:2006nq,Lorce:2007as,Lorce:2007fa, Lorce:2011kd, Diakonov:2005ib}, in light-cone constituent quark model\cite{Lorce:2011ni,Lorce:2011kd,Boffi:2002yy,Boffi:2003yj},  light-cone spectator model\cite{Liu:2015eqa},  light-cone quark-diquark model(LFQDM)\cite{Chakrabarti:2016yuw,Chakrabarti:2017teq}, light-front dressed quark model\cite{Mukherjee:2014nya, Mukherjee:2015aja,More:2017zqq}.

The generalized transverse-momentum-dependent distributions (GTMDs) are functions of the light-cone three-momenta of the parton and the momentum transfer to the hadron. The light-cone energy integrated Wigner distributions are interpreted as a Fourier transform of GTMDs. One of the intriguing aspects of GTMDs is their sensitivity to the skewness parameter $\xi$, which represents the longitudinal momentum transfer to the hadron. 
However, $\xi=0$ indicates that the momentum transfer occurs solely in the transverse direction.  
The GTMDs with non-zero skewness offer insights into the off-forward kinematics of parton distributions and provide a window into more complete three-dimensional tomography of the proton. 
Moreover, experimental observations involve 
$\xi \neq 0$ and demand a more comprehensive understanding of GTMDs with non-zero skewness. 
The quark GTMDs can be accessed through the physical exclusive double Drell–Yan process as proposed in~\cite{Bhattacharya:2017bvs}, while the gluon GTMDs are measurable in diffractive di-jet production in deep-inelastic lepton-nucleon and lepton-nucleus scattering~\cite{Hatta:2016dxp,Ji:2016jgn,Hatta:2016aoc,Bhattacharya:2022vvo} and ultra-peripheral proton-nucleus collisions~\cite{Hagiwara:2017fye}, as well as in virtual photon-nucleus quasi-elastic scattering~\cite{Zhou:2016rnt}. 
Recently, the detection of quark OAMs and gluon GTMDs signatures at the EIC from exclusive $\pi_0$ production and exclusive heavy (axial-)vector meson production are proposed in \cite{Bhattacharya:2026qnd,Bhattacharya:2023hbq}.
Under the bilinear decomposition of the GTMDs correlator, all polarization specific leading twist GTMDs, in principle, can have Time reversal odd (T-odd) part along with T-even part and be expressed as $\chi_{i,j} = \chi^{(e)}_{i,j} + i \chi^{(o)}_{i,j}$.  The T-even GTMDs are extensively studied for both quarks and gluons. Most prior model analysis has assumed momentum transfer only in the transverse direction(i.e., skewness $\xi=0$) and presented quark and gluon T-even GTMDs at leading twist  \cite{Chakrabarti:2016yuw,Chakrabarti:2017teq,Liu:2015eqa,Lorce:2011ni,Lorce:2011kd,Mukherjee:2014nya, Mukherjee:2015aja}. Recently, the T-even leading GTMDs with skewness are presented in the LFQDM model\cite{Maji:2022tog}, in the quark spectator diquark model \cite{Yang:2025neu} and in dressed quark model \cite{Ojha:2022fls,Jana:2023btd}.
However, 
some of the leading twist T-odd GTMDs are crucial as they are sensitive to the spin and momentum correlations within hadrons. The T-odd GTMDs provide insights into phenomena such as Sivers and Boer-Mulders Single-Spin Asymmetries (SSA)\cite{Sivers:1989cc,Boer:1997nt}, spin density of quarks, etc.
 Sivers effects have been measured in experiments by HERMES, COMPASS, and JLab \cite{COMPASS:2010hbb, COMPASS:2008isr, JeffersonLabHallA:2011ayy, HERMES:2009lmz} and investigated phenomenologically \cite{DAlesio:2004eso,Efremov:2004tp,Anselmino:2005ea,Collins:2005rq,Anselmino:2008sga,Anselmino:2013rya, Anselmino:2016uie,Martin:2017yms,Barone:2009hw,Zhang:2008nu}.
T-odd GTMDs and WDs distributions are not explored despite their connection to spin-momentum correlation. 
In this article, we consider the semi-inclusive deep inelastic scattering process (SIDIS) and extensively discuss such T-odd quark GTMDs $F^{(o)}_{1,2}$ and $H^{(o)}_{1,1}$ with non-zero skewness associated to Sivers and Boer-Mulders asymmetries. Another major focus of this paper is to investigate associated T-odd WDs in the impact parameter space as well as in the boost-invariant longitudinal position space.  

The six-dimensional Wigner distribution with three-momentum ${\bf p}$ and three-position ${\bf b }$  neglects the relativistic effect\cite{Ji:2003ak}.  For a relativistic strongly interacting bound state, an integrated five-dimensional Wigner distribution provides the density of finding a particle with three-momentum $(x,\bfp)$ and three-position $ \bfb $ \cite{Lorce:2011kd}. Recently, the six-dimensional WDs have been investigated adopting the spectator model LFWFs\cite{Yang:2025neu}.  In this work, we investigate the six-dimensional T-odd WDs in the LFQDM for a proton.

Recent advances in the study of WDs in the boost-invariant longitudinal space for the proton have gathered significant attention. The distributions in the boost invariant longitudinal space feature a long-distance tail as reported in Refs. \cite{Miller:2019ysh,Weller:2021wog}. 
The boost-invariant longitudinal position space is defined as $\sigma=\frac{1}{2}b^- P^+$, where $b^-$ is the longitudinal impact parameter and $P^+$ is the longitudinal momentum of the proton. The Fourier transform of GTMDs with respect to the skewness variable $\xi$ yields the Wigner distributions in the longitudinal position space $\sigma$. In light-front models framework, the $\sigma$-space have been investigated for other primary distributions e.g., the DVCS amplitudes and the GPDs of a relativistic spin-$\frac{1}{2}$ composite system \cite{Brodsky:2006ku,Chakrabarti:2008mw,Manohar:2010zm,Kumar:2015fta} in various models including those inspired by AdS/QCD \cite{Maji:2022tog,Ojha:2022fls} and phenomenological GPD models \cite{Brodsky:2006in,Chakrabarti:2008mw, Manohar:2010zm,Kumar:2015fta,Mondal:2015uha, Mondal:2017wbf,Chakrabarti:2015ama}.
A notable feature observed in longitudinal position space distribution is a diffraction pattern analogous to the diffractive scattering of a wave in optics. 

The Wigner distribution in impact parameter space $\bfb$ is obtained by taking the Fourier transform of GTMDs with respect to the transverse momentum transfer $\bfd$, more precisely, $\bfb$ is Fourier conjugate to $\bfD=\frac{\bfb}{(1-\xi^2)}$ for the case of non-zero skewness $\xi$ \cite{Diehl:2002he,Burkardt:2002hr,Ralston:2001xs,Kaur:2018ewq}. 
 It allows investigation of various partonic correlations, including spin-spin and spin-orbit interactions, and their dependence on the transverse position within the proton \cite{Chakrabarti:2017teq}. Transverse impact parameter space configuration of hadrons has been extensively explored ~\cite{Lorce:2011ni,Lorce:2011kd,Lorce:2011dv,Mukherjee:2014nya,Mukherjee:2015aja,More:2017zqq,Liu:2015eqa,Chakrabarti:2016yuw,Chakrabarti:2017teq,Chakrabarti:2019wjx,Gutsche:2016gcd,Kaur:2019lox,Kumar:2017xcm,Kanazawa:2014nha,Ma:2018ysi,Kaur:2019jow,Kaur:2019kpi,Zhang:2021tnr}.

This paper aims to explore leading-twist T-odd GTMDs of the proton with non-zero skewness using the extended light-front quark diquark model framework within the Dokshitzer Gribov Lipatov Altarelli Parisi (DGLAP) region. In this model, the T-odd sector is determined by incorporating the contribution from final state interaction (FSI), and the wave function of the model is constructed in the framework of the AdS/QCD soft-wall model. A brief overview of the T-odd sector in the Light-Front Quark Diquark Model(LFQDM) is given in the Sec.\ref{model}. The kinematics and definition of T-odd GTMD with skewness are included in Sec.\ref{sec_GTMD_def}, which contains analytical and numerical results of T-odd GTMDs in this model. In Sec.\ref{sec_spin_den}, the model results for quark-spin density are presented. In Sec.\ref{sec_WD_sig}, we present the results and discussions on boost-invariant longitudinal space for the Sivers and Boer-Mulders distributions. For completeness, the model results for Sivers and Boer-Mulder distributions in transverse impact parameter space are presented in Sec.\ref{sec_WD_bt}.

\section{T-odd sector in Light-front quark-diquark model \label{model}}
The Light-Front Quark Diquark Model (LFQDM) is a theoretical framework that simplifies the complex interactions of quarks and gluons by treating the proton as a bound state of a quark and a diquark \cite{Maji:2016yqo}.
In this model, both scalar and axial vector diquarks are considered, and the light-front wave functions (LFWFs) are derived from the two-particle effective wave functions obtained in the soft-wall AdS/QCD framework. It has been used successfully to describe various interesting properties of the nucleon, such as axial charge, tensor charge, PDFs, GPDs, TMDs, Wigner distributions at zero skewness, and spin asymmetries \cite{Maji:2015vsa, Maji:2016yqo,Maji:2017bcz,Maji:2017ill,Chakrabarti:2017teq,Maji:2017zbx,Maji:2017wwd,Kumar:2017dbf}. Recently, the T-even GTMDs at non-zero skewness for all polarization of proton are presented in\cite{Maji:2022tog}.

In LFQDM, considering $SU(4)$ spin-flavor symmetry, the  proton state $|P; \pm\rangle$ is written as superposition of the quark-diquark states as
$
|P; \pm\rangle = C_S|u~ S^0\rangle^\pm + C_V|u~ A^0\rangle^\pm + C_{VV}|d~ A^1\rangle^\pm\, \label{PS_state}
$
where $C_S, C_V$ and $C_{VV}$ are the coefficients of isoscalar-scalar diquark singlet state $|u~ S^0\rangle$, isoscalar-vector diquark state $|u~ A^0\rangle$  and isovector-vector diquark state $|d~ A^1\rangle$ respectively \cite{Jakob:1997wg,Bacchetta:2008af,Maji:2016yqo}. 
\begin{figure}[h]
\begin{center}
\includegraphics[scale=.4]{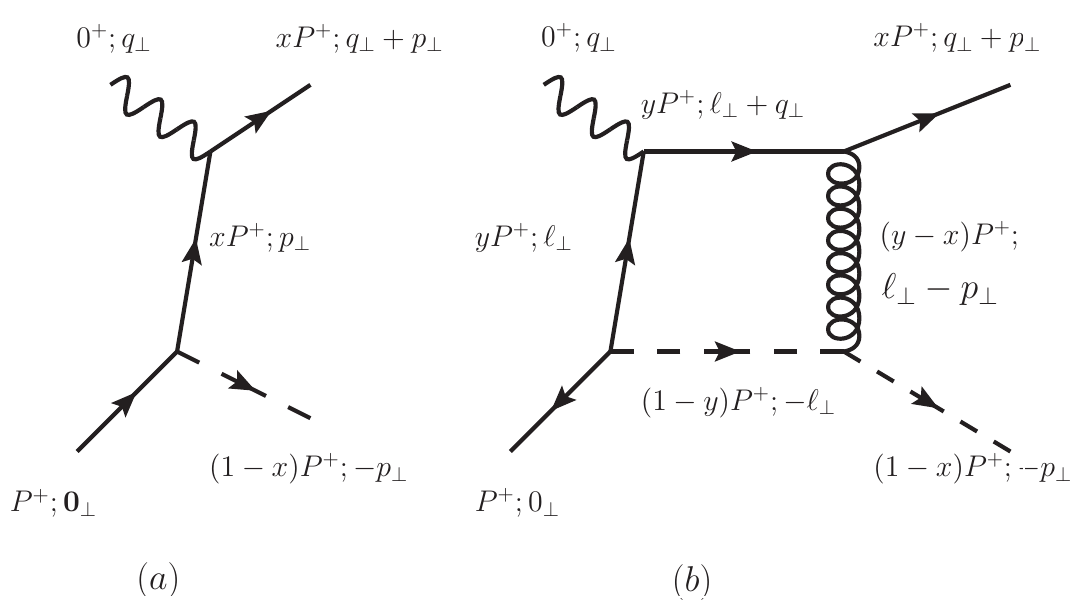} 
\caption{ SIDIS process at (a) tree level and (b)including final-state interaction (FSI)} \label{fig_FSI}
\end{center}
\end{figure}

For a scalar diquark, the two-particle Fock-state expansion with $J^z =\pm1/2$ is given by
\be
|u~ S\rangle^\pm & =& \int \frac{dx~ d^2\bfp}{2(2\pi)^3\sqrt{x(1-x)}} \sum_{\lambda_q=\pm 1/2} \psi^{\pm(u)}_{\lambda_q}(x,\bfp)~|\lambda_q~s; xP^+,\bfp\rangle \,,\label{fock_PS}
\ee
where $|\lambda_q~\lambda_S; xP^+,\bfp\rangle$ represents two-particle singlet state with scalar diquark  of helicity $\lambda_S=0$ . The light-front wave functions (LFWFs), with nucleon helicities $\lambda_N=\pm$ and for quark $\lambda_q=\pm$,  denoted by $\psi^{\lambda_N(u)}_{\lambda_q}(x,\bfp)$   have explicit form for different combinations of spin (plus and minus correspond to $+\frac{1}{2}$ and $-\frac{1}{2}$, respectively) as \cite{Lepage:1980fj, Maji:2017wwd, Ellis:2008in}
\be
\psi^{+(u)}_+(x,\bfp)&=&\psi^{-(u)}_-(x,\bfp) = N_S~\bigg[1+i \frac{e_1 e_2}{8 \pi}(\bfp^2 + B)g_1 \bigg] \varphi^{(u)}_{1}(x,\bfp),\nonumber \\
\psi^{+(u)}_-(x,\bfp)&=& N_S\bigg(- \frac{p^1+ip^2}{xM} \bigg) \bigg[1+i \frac{e_1 e_2}{8 \pi}(\bfp^2 + B)g_2\bigg]  \varphi^{(u)}_{2}(x,\bfp),\label{LFWF_S}\\
\psi^{-(u)}_+(x,\bfp)&=& N_S \bigg(\frac{p^1-ip^2}{xM}\bigg) \bigg[1+i \frac{e_1 e_2}{8 \pi}(\bfp^2 + B)g_2\bigg] \varphi^{(u)}_{2}(x,\bfp),\nonumber 
\ee

 Meanwhile, the state with spin-1 diquark is expressed as 
\be
|\nu~ A \rangle^\pm & =& \int \frac{dx~ d^2\bfp}{2(2\pi)^3\sqrt{x(1-x)}} \sum_{\lambda_q=\pm 1/2;} \sum_{\lambda_D=\pm 1, 0} \psi^{\pm(\nu)}_{\lambda_q \lambda_D}(x,\bfp)~|\lambda_q~ \lambda_D; xP^+,\bfp\rangle \,,\label{fock_PA}
\ee
where $|\lambda_q~\lambda_D; xP^+,\bfp\rangle$ corresponds to the two-particle state with the axial-vector diquark of helicities $\lambda_D=\pm 1,0$~(triplet).
While, the wavefunctions $\psi^{\lambda_N(u)}_{\lambda_q, \lambda_D}(x,\bfp)$  
for axial-vector diquark (for $J= \pm 1/2$ ) carries an additional subscript corresponding to the non-singlet diquark helicities, and the explicit form for different combinations is given by \cite{Maji:2017wwd}
\be
\psi^{+(\nu)}_{+~+}(x,\bfp)=& \frac{\left(p^1 - ip^2\right)^2}{\bfp^2} ~\psi^{-(\nu)}_{-~-}(x,\bfp) &= N^{(\nu)}_1 \sqrt{\frac{2}{3}} \bigg(\frac{p^1-ip^2}{xM}\bigg) \bigg[1+i \frac{e_1 e_2}{8 \pi}(\bfp^2 + B)g_2 \bigg] \varphi^{(\nu)}_{2}(x,\bfp),\nonumber \\
\psi^{+(\nu)}_{-~+}(x,\bfp)=& - ~\psi^{-(\nu)}_{+~-}(x,\bfp) &= N^{(\nu)}_1 \sqrt{\frac{2}{3}}  \bigg[1+i \frac{e_1 e_2}{8 \pi}(\bfp^2 + B)g_1 \bigg] \varphi^{(\nu)}_{1}(x,\bfp),\nonumber\\
\psi^{+(\nu)}_{+~0}(x,\bfp)=& - ~\psi^{-(\nu)}_{-~0}(x,\bfp) &= - N^{(\nu)}_0 \sqrt{\frac{1}{3}} \bigg[1+i \frac{e_1 e_2}{8 \pi}(\bfp^2 + B)g_1 \bigg] \varphi^{(\nu)}_{1}(x,\bfp),\label{LFWF_Vp}\\
\psi^{+(\nu)}_{-~0}(x,\bfp)=& \frac{\left(p^1+ip^2\right)^2}{\bfp^2} ~\psi^{-(\nu)}_{+~0}(x,\bfp) &= N^{(\nu)}_0 \sqrt{\frac{1}{3}} \bigg(\frac{p^1+ip^2}{xM} \bigg)  \bigg[1+i \frac{e_1 e_2}{8 \pi}(\bfp^2 + B)g_2 \bigg] \varphi^{(\nu)}_{2}(x,\bfp),\nonumber 
\ee
\be
\psi^{+(\nu)}_{+~-}(x,\bfp)=& \psi^{+(\nu)}_{-~-}(x,\bfp)= \psi^{-(\nu)}_{+~+}(x,\bfp)&=\
\psi^{-(\nu)}_{-~+}(x,\bfp)= 0,\nonumber 
\ee
with normalization constants $N_S, N^{(\nu)}_0$ and $N^{(\nu)}_1$ for the respective flavour $\nu=u,d $. The wave functions are normalized according to the quark counting rules \cite{Maji:2016yqo}.
The imaginary part of the wave functions is consistent with the final state interaction (FSI) amplitude \cite{Brodsky:2002cx}. In the imaginary part, term $g_1, g_2$ are given by 
\be
g_1 &=& \int^1_0 d\alpha \frac{-1}{\alpha(1-\alpha)\bfp^2 + \alpha m_g^2 +
(1-\alpha)B}, \label{g1}\\
g_2 &=& \int^1_0 d\alpha \frac{-\alpha}{\alpha(1-\alpha)\bfp^2 + \alpha m_g^2 +
(1-\alpha)B},\label{g2}
\ee
 and $\bfp$ independent term $B$ has explicit form
 \be 
B &=& x(1-x)(-M^2+\frac{m^2_q}{x}+\frac{m^2_D}{1-x}),\label{B}
\ee
with proton mass $M$, struck quark and diquark mass $m_q$ and $m_D$, and gluon mass $ m_g$. The gluon mass is taken as zero at the end of the calculations. Here, the final state interaction is governed by a gluon, and $e_1, e_2$ are the color charges of the struck quark and diquark, respectively. Thus, FSI gauge exchange strength is considered as $\frac{e_1 e_2}{4 \pi} \rightarrow - C_F \alpha_s $ with color factor $C_F$. The LFWFs $\varphi^{(\nu)}_i(x,\bfp)$ are the modified form of the soft-wall AdS/QCD prediction, and the two-particle effective wave functions read as
\be
\varphi_i^{(\nu)}(x,\bfp)=\frac{4\pi}{\kappa}\sqrt{\frac{\log(1/x)}{1-x}}x^{a_i^\nu}(1-x)^{b_i^\nu}\exp\bigg[-\frac{\bfp^2}{2\kappa^2}\frac{\log(1/x)}{(1-x)^2}\bigg]\,,
\label{LFWF_phi}
\ee
with flavour index $\nu=u, d$. The modification is incorporated through the parameters $a_i^\nu, b_i^\nu$ and $\varphi_i^\nu ~(i=1,2)$ of  Eq.(\ref{LFWF_phi}) reduces to the original AdS/QCD wave function \cite{Brodsky:2007hb,deTeramond:2012rt} at the limit $a_i^\nu=b_i^\nu=0$. We use the AdS/QCD scale parameter $\kappa =0.4$ GeV~\cite{Chakrabarti:2013dda,Chakrabarti:2013gra} and the quarks are assumed to be massless. The parameters of this model are obtained by fitting the flavor-decomposed Dirac and Pauli form factors data, as detailed in Refs.~\cite{Maji:2016yqo, Maji:2017bcz}. The wave functions generated by this model, using these parameters, show a reasonably good agreement with the proton electric and magnetic charge radius data, axial charge, and tensor charge as well as parton distribution measurements.

%

\section{T-odd GTMDs with non-zero skewness}\label{sec_GTMD_def}

In deep inelastic scattering, a proton of average momentum $P$ interacts with a lepton via a photon of virtuality $Q^2=-q^2$. Eventually, the high-energy virtual photon interacts with the interior parton of momentum $p$, which caries longitudinal momentum fraction $x=p^+/P^+$ and transverse momentum $\bfp$. The momentum transfer to the system is defined by $\Delta$ and $-t = \Delta^2$. 
For such interaction, the GTMDs correlator is defined at the fixed light cone time $ z^+=0$ as~\cite{Meissner:2008ay,Meissner:2009ww}
\be
W^{\nu [\Gamma]}_{[\lambda^{\prime\prime}\lambda^{\prime}]}(x,\xi,\bfd^2, \bfp^2, \bfp.\bfd)=\frac{1}{2}\int \frac{dz^-}{(2\pi)} \frac{d^2z_T}{(2\pi)^2} e^{ip.z} 
\langle P^{\prime\prime}; \lambda^{\prime\prime} |\bar{\psi}^\nu (-z/2)\Gamma \mathcal{W}_{[-z/2,z/2]} \psi^\nu (z/2) |P^\prime;\lambda^{\prime}\rangle \bigg|_{z^+=0}\,,
\label{Wdef}
\ee
where $|P^\prime;\lambda^{\prime}\rangle $ and $|P^{\prime\prime}; \lambda^{\prime\prime}\rangle$ are the initial and final states of the proton with helicities $\lambda^\prime$ and $\lambda^{\prime\prime}$, respectively. The bi-local operator with quark fields $\psi $ and $\bar{\psi}$ is at two different points linked by the gauge-link   $\mathcal{W}_{[-z/2,z/2]}$, which is chosen to be unity. 
 Dirac $\gamma$-matrices, i.e., $\Gamma=\{\gamma^+,\, \gamma^+\gamma^5,\, i\sigma^{j+} \gamma^5\}$ are chosen to extract the contribution of unpolarized, longitudinally polarized and transversely polarized quarks, respectively. 
In the symmetric frame, the average four-momentum of proton $P= \frac{1}{2} (P^{\prime\prime}+P^\prime)$ can be written as 
\be 
P &\equiv& \bigg(P^+,\frac{M^2+\bfd^2/4}{(1-\xi^2)P^+},\textbf{0}_\perp\bigg)\,,
\ee
where, the initial ($P^\prime$) and final ($P^{\prime\prime}$) momentum of proton are given by
\be
P^{\prime} &\equiv& \bigg((1+\xi)P^+,\frac{M^2+\bfd^2/4}{(1+\xi)P^+},-\bfd/2\bigg)\,,\label{Pp}\\
P^{\prime\prime} &\equiv& \bigg((1-\xi)P^+,\frac{M^2+\bfd^2/4}{(1-\xi)P^+},\bfd/2\bigg)\,. \label{Ppp}
\ee
The interior parton has non-zero transverse momentum $p \equiv \bigg(xP^+, p^-,\bfp \bigg)$, and the sum of transverse momenta of all the constituent partons vanishes, which satisfies the zero transverse momentum of the proton $e.g.,$ $P_\perp = \sum_i \bfp^i = 0$.   
In a symmetric frame, the energy transferred to the system $\Delta=(P^{\prime\prime}-P^\prime)$ reads as
\be 
\Delta &\equiv& \left(-2\xi P^+, \frac{t + \bfd^2}{-2 \xi P^+},\bfd \right)\,,
\ee
where, skewness is defined as $\xi=- \Delta^+/2P^+$. The energy transferred to the process is measured as the square of total momentum transfer $t= \Delta^2$, and an explicit relation can be established using $\Delta^-=( P^{\prime \prime -} - P^{\prime  -})$ as
 \be 
- t= \frac{4 \xi^2 M^2 + \bfd^2}{(1-\xi^2)}\,. \label{mt_def}
\ee 
Here, the definition of skewness $\xi$ is adopted from the convention in Ref.~\cite{Meissner:2009ww}, which differs by a minus sign with respect to the definition in Ref.~\cite{Brodsky:2000xy}.

For a spin-$1/2$ hadron, the bilinear decomposition of the fully unintegrated quark-quark correlator is parameterized in terms of GTMDs as \cite{Meissner:2009ww}. 
\be 
W^{\nu [\gamma^+]}_{[\lambda^{\prime\prime}\lambda^{\prime}]} (x,\xi,\bfd^2, \bfp^2, \bfp.\bfd) &=&\frac{1}{2M} \bar{u}(P^{\prime\prime},\lambda^{\prime\prime})\bigg[ \dots + \frac{i\sigma^{i+} \pp^i}{P^+}F_{1,2}(x,\xi,\bfd^2, \bfp^2, \bfp.\bfd) +  \dots \bigg] u(P^\prime,\lambda^\prime)\label{WV_def}\,,\\
W^{\nu [i \sigma^{j+}\gamma^5]}_{[\lambda^{\prime\prime}\lambda^{\prime}]} (x,\xi,\bfd^2, \bfp^2, \bfp.\bfd) &=&\frac{1}{2M} \bar{u}(P^{\prime\prime},\lambda^{\prime\prime})\bigg[-\frac{i\epsilon^{ij}_\perp \pp^i }{M} H_{1,1}(x,\xi,\bfd^2, \bfp^2, \bfp.\bfd) + \dots \bigg] u(P^\prime,\lambda^\prime)\,, \label{WT_def}
\ee
with the four component Dirac-spinor $u(k,\lambda) $ of momentum $k$ and the helicity $\lambda\,(=\pm )$. The explicit form of the Dirac spinors is appended in App.\ref{AppA}. We concentrate on the unpolarised and transversely polarised partons and corresponding Dirac structures $\Gamma =\gamma^+, i \sigma^{j+}\gamma^5 $ respectively, as these two encode T-odd distributions related to Sivers and Boer-Mulders distributions. The dots in the right-hand side of Eqs.(\ref{WV_def},\ref{WT_def}) indicate other leading twist GTMDs having different bilinear pre-factors whose complete discussion is given in \cite{Meissner:2009ww}. All the GTMDs can be written in terms of time reversal even (T-even) and time reversal odd (T-odd) parts as 
\be 
\mathcal{X}_{i,j}(x,\xi,\bfd^2, \bfp^2, \bfp.\bfd)= \mathcal{X}^e_{i,j}(x,\xi,\bfd^2, \bfp^2, \bfp.\bfd) + i \mathcal{X}^o_{i,j}(x,\xi,\bfd^2, \bfp^2, \bfp.\bfd)
\ee
where $\mathcal{X}$ stands for any GTMDs at leading twist. 
Using the two-particle Fock-state expansion of Eqs.(\ref{fock_PS},\ref{fock_PA}) in Eq.(\ref{Wdef}),  the GTMDs correlators can be expressed in terms of overlap representation as
\be 
 W^{ [\gamma^+]}_{[\lambda^{\prime\prime} \lambda^{\prime}]}(x,\xi,\bfd^2, \bfp^2, \bfp.\bfd)&=& C^2_S\frac{1}{16\pi^3} \sum_{\lambda_q} \psi^{\lambda^{\prime\prime}\dagger}_{\lambda_q}(x^{\prime\prime},\bfp^{\prime\prime})\psi^{\lambda^{\prime}\dagger}_{\lambda_q}(x^{\prime},\bfp^{\prime})\nonumber \\
&& \hspace{1cm} + C^2_A \frac{1}{16\pi^3} \sum_{\lambda_q}\sum_{\lambda_D} \psi^{\lambda^{\prime\prime}\dagger}_{\lambda_q\lambda_D}(x^{\prime\prime},\bfp^{\prime\prime})\psi^{\lambda^{\prime}\dagger}_{\lambda_q\lambda_D}(x^{\prime},\bfp^{\prime})\label{WpsiU}\\
W^{ [i\sigma^{j+}\gamma^5]}_{[\lambda^{\prime\prime} \lambda^{\prime}]}(x,\xi,\bfd^2, \bfp^2, \bfp.\bfd)&=& C^2_S \frac{1}{16\pi^3} \sum_{\lambda^{\prime\prime}_q}\sum_{\lambda^\prime_q} (2 \lambda^\prime_q i)^i~ \psi^{\lambda^{\prime\prime}\dagger}_{\lambda^{\prime\prime}_q}(x^{\prime\prime},\bfp^{\prime\prime})\psi^{\lambda^{\prime}\dagger}_{\lambda^\prime_q}(x^{\prime},\bfp^{\prime})\nonumber \\
&& \hspace{.5cm} + C^2_A \frac{1}{16\pi^3} \sum_{\lambda^{\prime\prime}_q}\sum_{\lambda^{\prime}_q}\sum_{\lambda_D} \epsilon^{ij}_\perp (2\lambda^\prime_q i)^i  ~ \psi^{\lambda^{\prime\prime}\dagger}_{\lambda^{\prime\prime}_q\lambda_D}(x^{\prime\prime},\bfp^{\prime\prime})\psi^{\lambda^{\prime}\dagger}_{\lambda^{\prime}_q\lambda_D}(x^{\prime},\bfp^{\prime})\label{WpsiT}
\ee
where, $C_A=C_V, C_{VV}$ for $u$ and $d$ quarks respectively \cite{Maji:2016yqo}.
 The explicit form of the initial and final transverse momenta of the struck quark are given by
\be 
\bfp^{\prime}=\bfp-(1-x^\prime)\frac{\bfd}{2},& \quad & {\rm with} \quad x^{\prime}=\frac{x+\xi}{1+\xi}\,,\label{ptp}\\ 
\bfp^{\prime\prime}=\bfp+(1-x^{\prime\prime})\frac{\bfd}{2}\,, & \quad &  {\rm with} \quad  x^{\prime\prime}=\frac{x-\xi}{1-\xi}\,, \label{ptpp}
\ee
respectively. 

\subsection{Model Results for GTMDs}
In this LFQDM, the GTMD-correlator is written in terms of the scalar and the axial-vector diquark components as
\be 
W^{\nu[\Gamma]}_{[\lambda^{\prime\prime}\lambda^\prime]}(x, \bfp, \bfd) &=& C^2_S ~W^{\nu[\Gamma](S)}_{[\lambda^{\prime\prime}\lambda^\prime]}(x, \bfp, \bfd) +  C^2_A ~W^{\nu[\Gamma](A)}_{[\lambda^{\prime\prime}\lambda^\prime]}(x, \bfp, \bfd)\,,
\ee
where, $C_A=C_V, C_{VV}$ for $u$ and $d$ quarks respectively.

In this work, our main concern is to address $F^o_{1,2}(x,\xi,\bfd^2, \bfp^2, \bfp.\bfd)$ and $ H^o_{1,1}(x,\xi,\bfd^2, \bfp^2, \bfp.\bfd)$ which reduced to the Sivers and Boer-Mulders functions respectively at the TMDs limit $\xi=0, \bfd=0$. 
From the overlap representation of Eqs.(\ref{WpsiU},\ref{WpsiT}) and the bilinear decomposition of Eqs.(\ref{WV_def},\ref{WT_def}), the final results of the odd part of the GTMDs in this model read as
\be 
F^{(o)\nu}_{1,2}(x,\xi,\bfd^2,\bfp^2,\bfd.\bfp) &=& \frac{1}{16\pi^3}\bigg[ N^\nu_{F12}\frac{1}{\zf} \bigg\{\left( \chi^{\prime\prime}_1 - \chi^\prime_2 \right) \frac{\Aodp \Atp}{x^\prime} - \left(\chi^{\prime\prime}_2 - \chi^\prime_1\right) \frac{\Atdp \Aop}{x^{\prime\prime}} \bigg\} \nonumber \\
&& - \frac{\bfd^2}{2M^2} N^\nu_{F14} \frac{\xi}{(1-\xi^2)^{3/2}} \frac{(1-x)}{x^{\prime\prime}x^\prime} \left( \chi^{\prime\prime}_2 - \chi^\prime_2 \right) \Atdp\Atp \bigg] \exf\,,\nonumber\\ \label{F12}\\    
H^{(o)\nu}_{1,1}(x,\xi,\bfd^2,\bfp^2,\bfd.\bfp) &=&  N^\nu_{H11}\frac{1}{16\pi^3}  \zf \bigg[  \left(\chi^{\prime\prime}_1 - \chi^\prime_2\right)\frac{\Aodp\Atp}{x^\prime} -  \left(\chi^{\prime\prime}_2 -\chi^\prime_1 \right) \frac{\Atdp\Aop}{x^{\prime \prime}} \bigg] \nonumber \\
&& \hspace{7cm} \times \exf\,,\label{H11}
\ee

Where,
\be 
 A^\nu_i(x) &=& \frac{4\pi}{\kappa}\sqrt{\frac{\log(1/x)}{1-x}}x^{a_i^\nu}(1-x)^{b_i^\nu},\\
\tilde{a}(x)&=& \frac{\log(1/x)}{2\kappa^2 (1-x)^2};~~~{\rm and}~~
a(x)= 2 \tilde{a}(x).
\ee
The contribution from the imaginary part of the wave function is denoted by
\be 
\left(\chi^{\prime\prime}_1 - \chi^\prime_2 \right) &=& \frac{1}{2} C_F \alpha_S \left[ \ln \left(1+ \frac{\bfp^{\prime\prime 2}}{B(x^{\prime \prime})}\right) + \frac{B(x^\prime)}{\bfp^{\prime 2}} \ln \left(1+ \frac{\bfp^{\prime 2}}{B(x^\prime)}\right)\right], \\
\left(\chi^{\prime\prime}_2 - \chi^\prime_1 \right) &=& -\frac{1}{2} C_F \alpha_S \left[ \frac{B(x^{\prime \prime})}{\bfp^{\prime\prime 2}} \ln \left(1+ \frac{\bfp^{\prime\prime 2}}{B(x^{\prime \prime})}\right) + \ln\left(1+ \frac{\bfp^{\prime 2}}{B(x^\prime)}\right) \right], \\
\left(\chi^{\prime\prime}_2 - \chi^\prime_2 \right) &=& -\frac{1}{2} C_F \alpha_S \left[\frac{B(x^{\prime \prime})}{\bfp^{\prime\prime 2}} \ln \left(1+ \frac{\bfp^{\prime\prime 2}}{B(x^{\prime \prime})}\right) - \frac{B(x^\prime)}{\bfp^{\prime 2}} \ln\left(1+ \frac{\bfp^{\prime 2}}{B(x^\prime)}\right) \right].
\ee
\begin{figure}[ht]
\includegraphics[scale=.2]{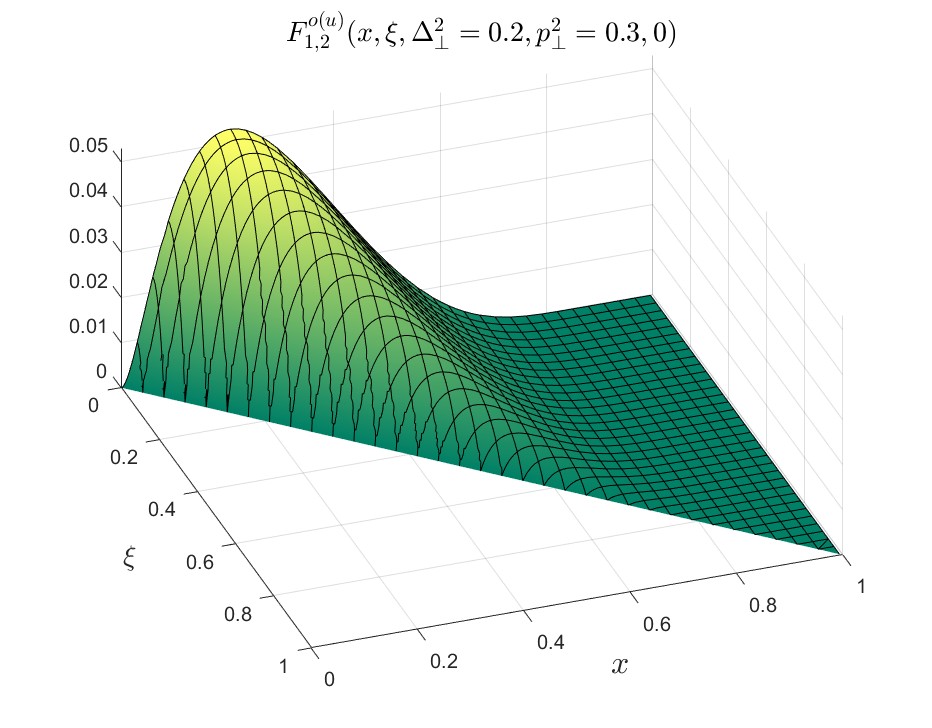} 
\includegraphics[scale=.2]{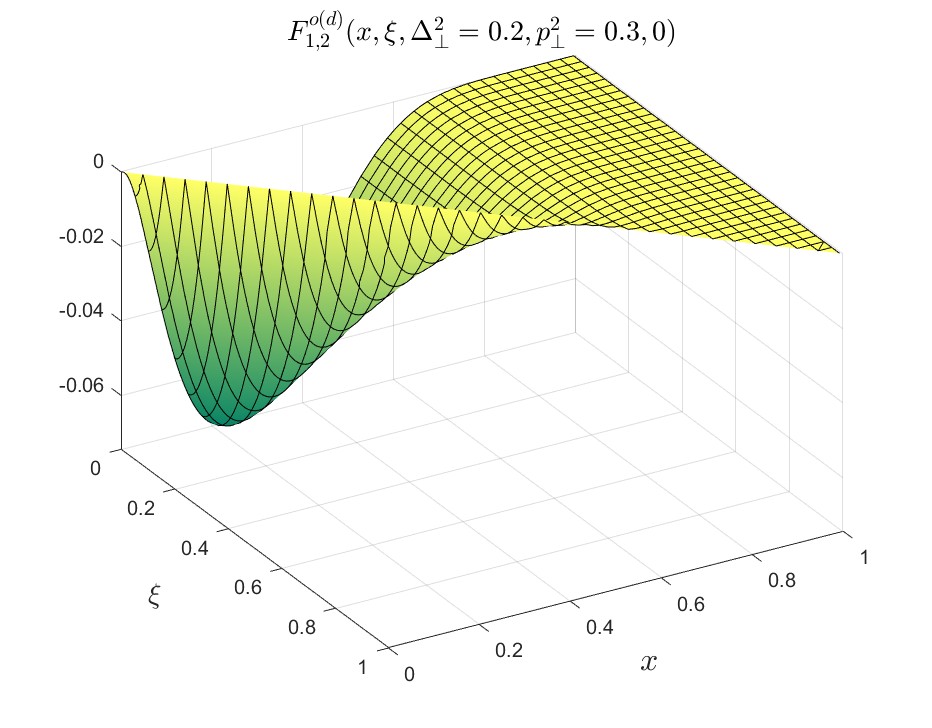}
\includegraphics[scale=.2]{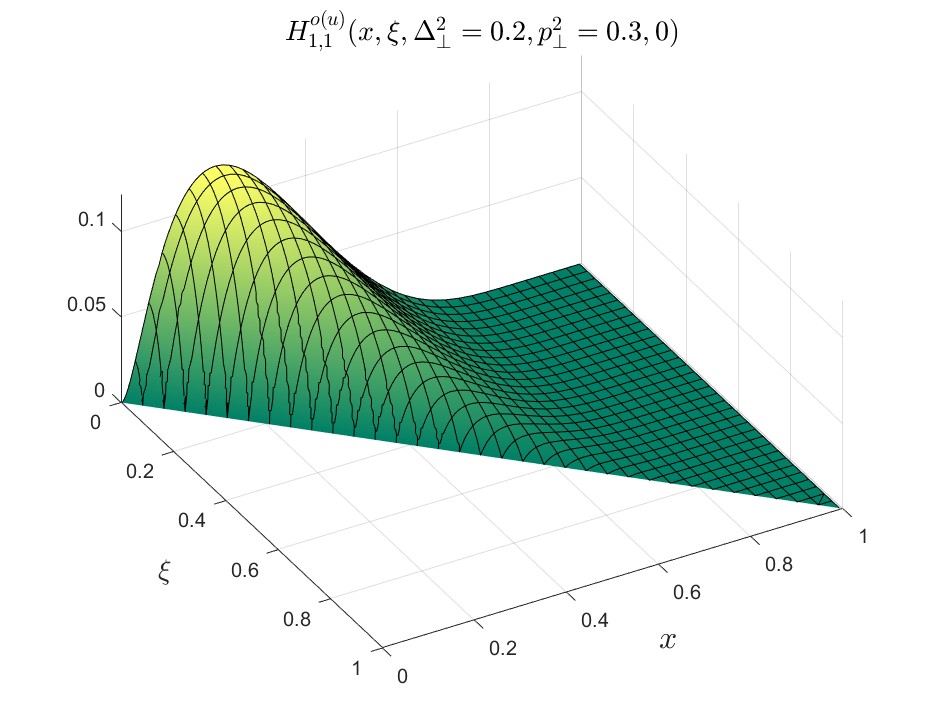} 
\includegraphics[scale=.2]{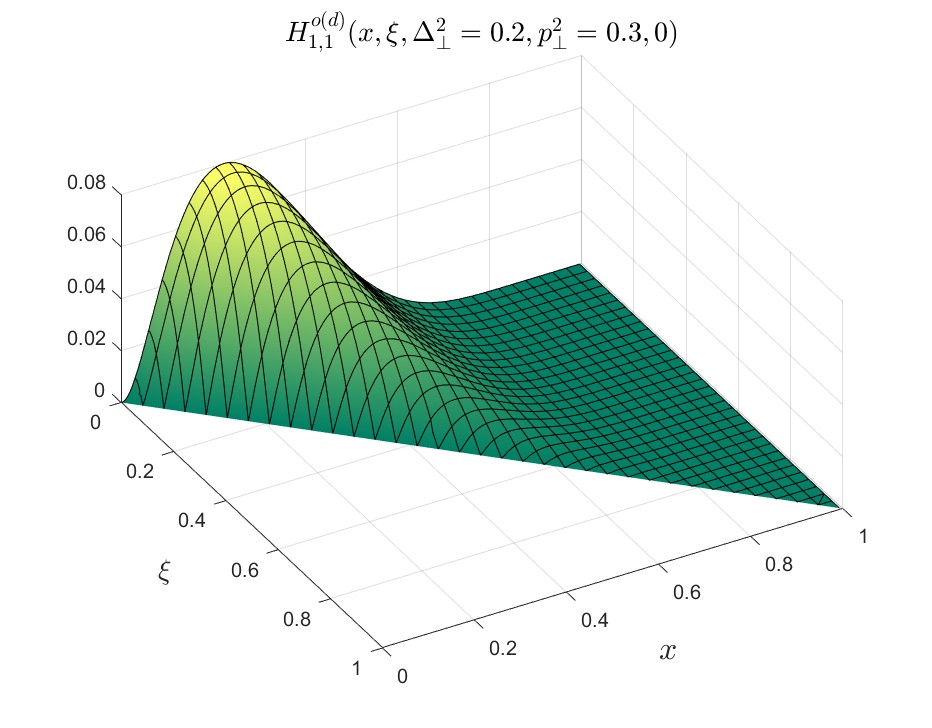}
\caption{Model results of $F^{(0)}_{1,2}$ (upper row) and $H^{(0)}_{1,1}$ (lower row) are shown in $\xi$-$x$ plane for the DGLAP region $\xi < x < 1$. The left and right columns are for $u$ and $d$ quarks respectively with $\bfp^2=0.3~GeV^2$ and  $\bfd^2=0.2~GeV^2$.}\label{fig_GTMD_xz}
\end{figure} 
These combinations cancel out the divergent part of the integration function $g_1$ and $g_2$ and produce a finite contribution. At $\xi=0$ limit, the difference in longitudinal fractions of incoming and outgoing partons vanishes $x^\prime=x^{\prime\prime}=x$ and the exponential term becomes independent of $\bfp.\bfd$. The third term of Eq.(\ref{F12}) vanishes at the zero-skewness. The normalization constants of the distributions are denoted by  
$
N^\nu_{F12} =  \big(C^2_SN^2_S-C^2_A \frac{1}{3}N^2_0\big)^\nu\,, ~
N^\nu_{F14}
= \big(C^2_SN^2_S+C^2_A\big(\frac{1}{3}N^2_0-\frac{2}{3}N^2_1\big)\big)^\nu\,,~ 
 N^\nu_{H11}
=  \big(C^2_SN^2_S+C^2_A \big(\frac{1}{3}N^2_0+\frac{2}{3}N^2_1\big)\big)^\nu\,.
$
For completeness, the LFQDM results for all the other leading twist T-odd GTMDs are listed up in App.\ref{AppA}.

Here, we present model results of the T-odd GTMDs numerically and concentrate on the skewness variation. The model parameters are taken from Refs.\cite{Maji:2017bcz,Maji:2016yqo}. with proton mass $M=0.94~GeV$, diquark mass $m_D=0.98\pm 0.04~GeV$. In principle the mass of the diquark has a lower bound given by the proton mass to satisfy the condition of a stable bound-state proton. However, the upper bound of $m_D$ is restrict by the spin sum rule \cite{Chakrabarti:2024hwx}. The quarks are considered to be massless. 

Fig.\ref{fig_GTMD_xz} represents the illustration of flavor dependent T-odd GTMDs $F^{(0)}_{1,2}$ and $H^{(0)}_{1,1}$ in three-dimension of longitudinal momentum fraction $x$ and skewness $\xi$. We consider the DGLAP region where the longitudinal momentum fraction is restricted between skewness $\xi$ and its upper bound $1$ i.e., $\xi < x < 1$. 
 The upper row is for $F^{(0)}_{1,2}$  and the lower row is for $H^{(0)}_{1,1}$ distributions, while the left column shows for $u$-quark and right row is for  $d$-quark. All these numerical results are computed for $\bfp^2=0.3~GeV^2$ and  $\bfd^2=0.2~GeV^2$. We also consider the case when $\bfd$ is perpendicular to $\bfp$ i.e, $\bfp.\bfd=0$. 
The GTMD, $F^{(0)}_{1,2}$, is the distribution associated to an unpolarised parton in a transversely polarised proton. 
The model result of $F^{(0)}_{1,2}$ in the DGLAP region shows a positive peak for $u$-quark and a negative peak for $d$-quark, and the peak belongs to the lower value of $x<0.5$. The flip of polarity in $F^{(0)}_{1,2}$ with the change of flavor from  $u$ to $d$ quark is the consequence of Spin and transverse momentum correlation that led to the Sivers Effect at TMD limits. Interestingly, a similar left-right shifting is noticed for the more generic GTMDs distributions. However, the magnitude of the shift varies with skewness $\xi$.
More precisely, for $u$ quark, the peak of the distribution shifts towards the right as the momentum transfer in the longitudinal direction increases, while their magnitudes decrease. A similar right shift of the peak is noticed for  $d$-quarks with the opposite polarity peak. 
The GTMD $H^{(0)}_{1,1}$ provides a distribution of transversely polarised quarks in an unpolarized proton. 
The distributions $H^{(0)}_{1,1}$ in $\xi$-$x$ plain show positive peaks for both $u$ and $d$ quarks. The peaks shifted towards the right upto $x<0.5$ and the magnitude dies out with the increase of $\xi$. While the intensity of the peak is higher for the $u$-quark than for the $d$-quark.  

\begin{figure}[htb]
\includegraphics[scale=.2]{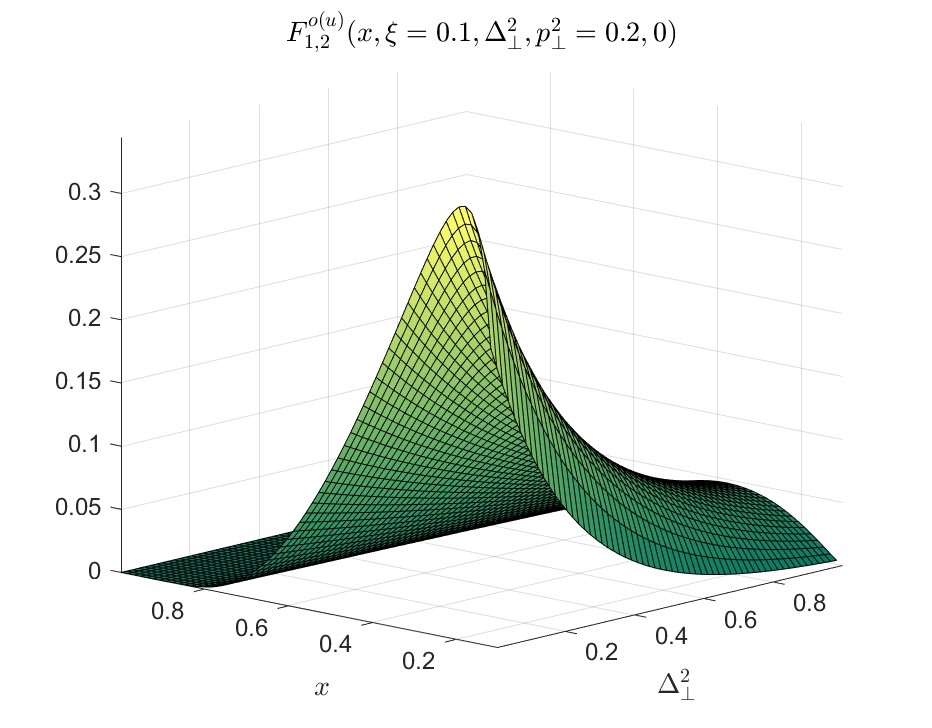} 
\includegraphics[scale=.2]{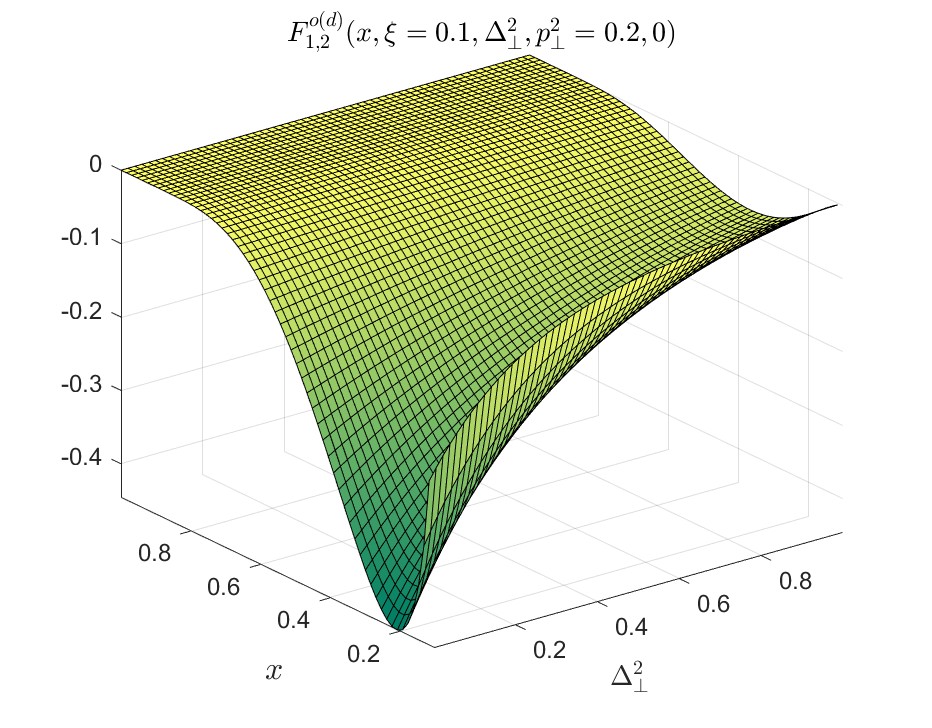}
\includegraphics[scale=.2]{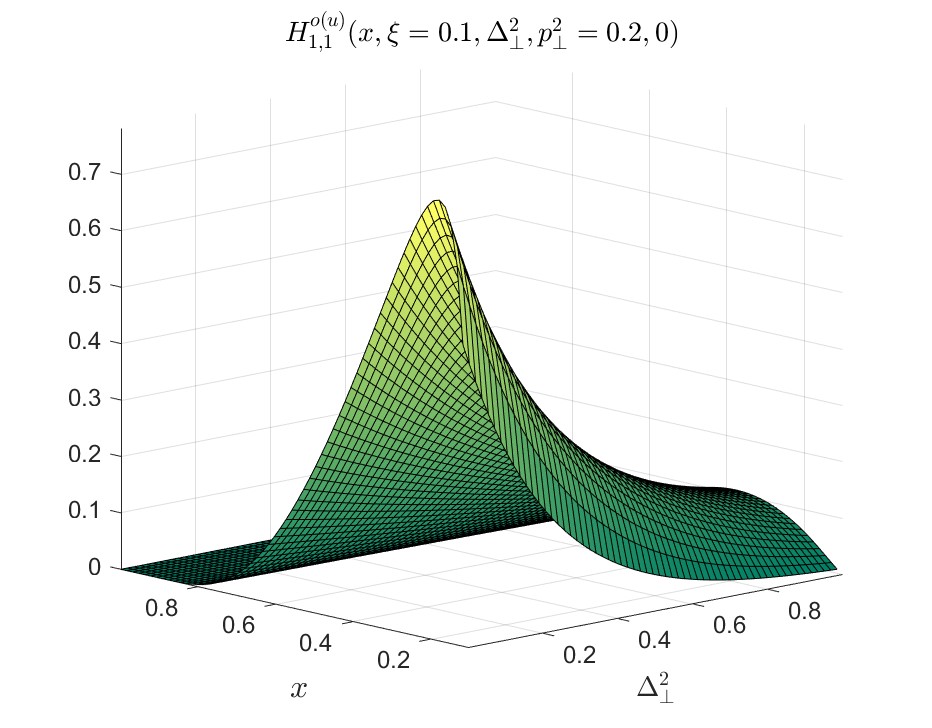} 
\includegraphics[scale=.2]{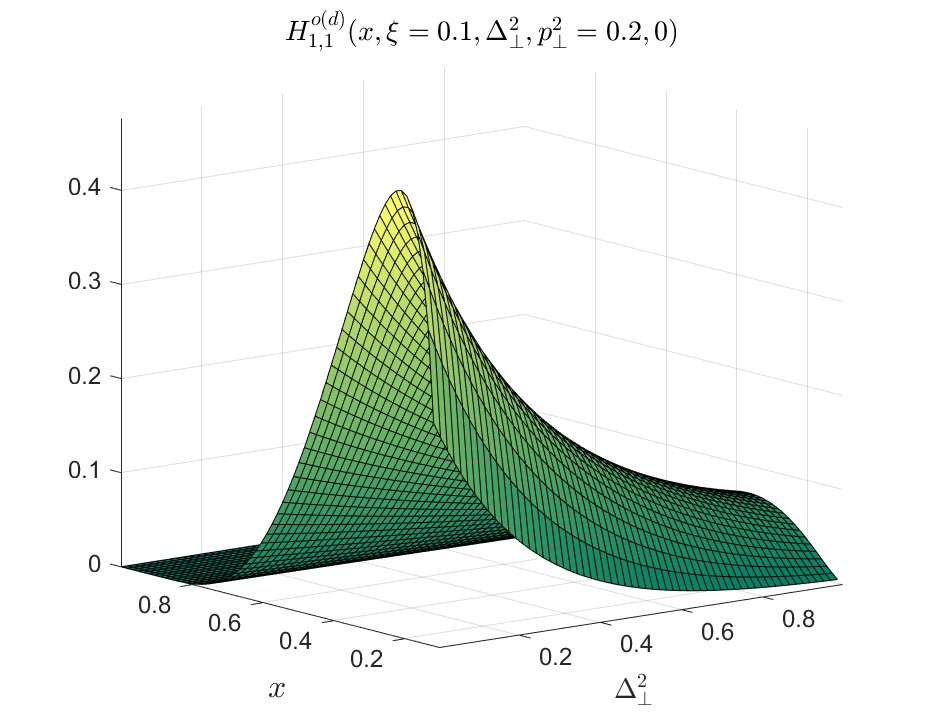}
\caption{Model results of $F^{(0)}_{1,2}$ (upper row) and $H^{(0)}_{1,1}$ (lower row) are shown in $x$-$\bfd^2$ plane for the DGLAP region $\xi < x < 1$. The left and right columns are for $u$ and $d$ quarks, respectively, with  $\xi=0.1$ and $\bfp^2=0.2~GeV^2$. }\label{fig_GTMD_xdt}
\end{figure}

Fig.\ref{fig_GTMD_xdt} describes the $x$ and $\bfd^2$ dependence of Sivers (upper row) and Boer-Mulders (lower row) GTMDs at fixed $\xi=0.1$ and $\bfp^2=0.2$ GeV$^2$. Thus, in the DGLAP region, the accessible space of $x$ is $0.1< x <1$. The left and right columns are for $u$ and $d$ quarks, respectively. 
The flavour decomposed results for Sivers and Boer-Mulders GTMDs, in $x$ and $\bfd^2$, show almost similar general feature-- the peaks of the distribution move towards larger values of $x$ with increasing momentum transfer $\bfd^2$, and the magnitudes of all the distributions decrease along $x$. To preserve the limited kinetic energy, the distributions in the transverse momentum broaden at higher-$x$, which reflects a trend
to carry a larger portion of the kinetic energy. Such general features are shown nearly model-independent properties for the GPDs and observed in several theoretical studies of the GPDs~\cite{Ji:1997gm,Scopetta:2002xq,Petrov:1998kf,Penttinen:1999th,Boffi:2002yy,Vega:2010ns,Chakrabarti:2013gra,Mondal:2015uha,Chakrabarti:2015ama,Mondal:2017wbf,deTeramond:2018ecg,Xu:2021wwj}. 
In this model, $F_{1,2}$ distribution flips polarity from positive to negative for $u$ and $d$, respectively. While $ H_{1,1}$ distributions remains positive irrespective of quark flavours. The different polarities of $F_{1,2}$ for the $u$ and $d$ quarks lead to the Sivers effect \cite{Sivers:1989cc},
 which indicates an unpolarised quark distribution in a transversely polarized target has the transverse momentum asymmetry in the perpendicular direction to the proton spin.
 

%
 
\subsection{Spin-Momentum Correlation and Spin Density \label{sec_spin_den}}
\begin{figure}[h]
\includegraphics[scale=.15]{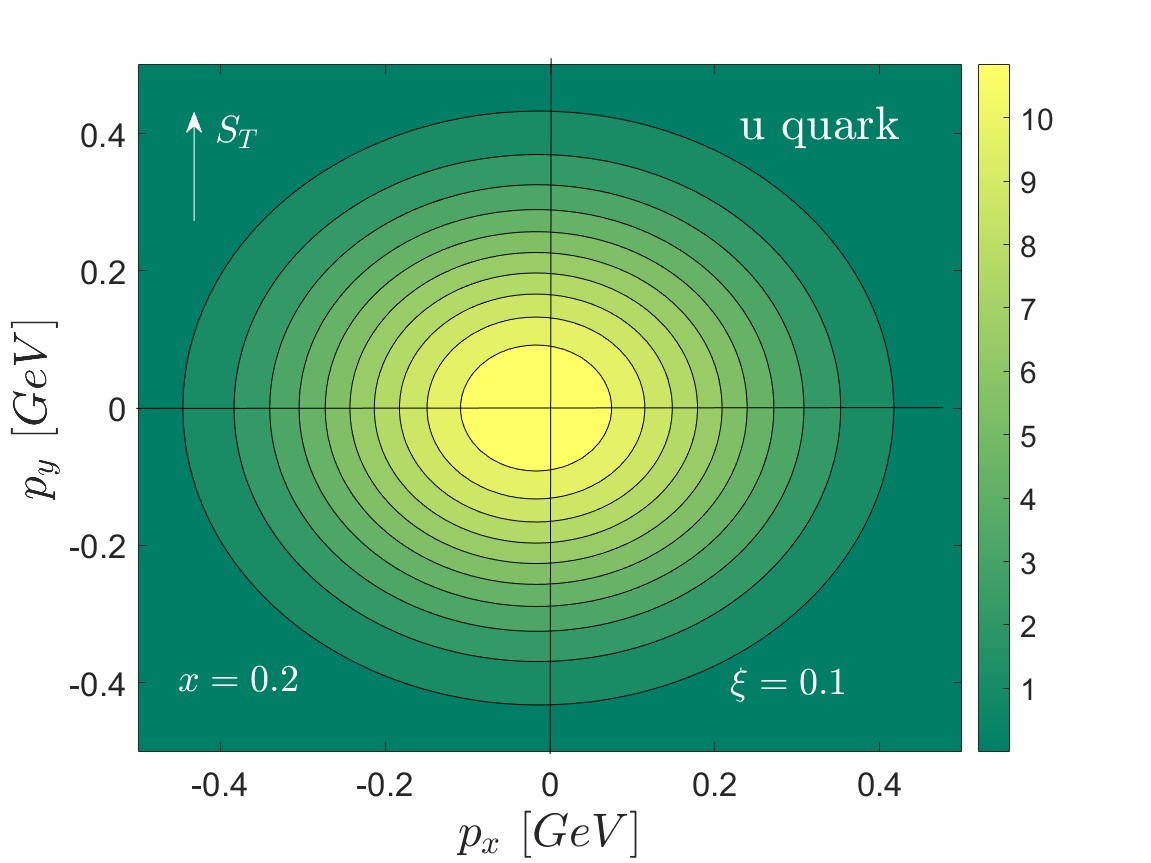} 
\includegraphics[scale=.15]{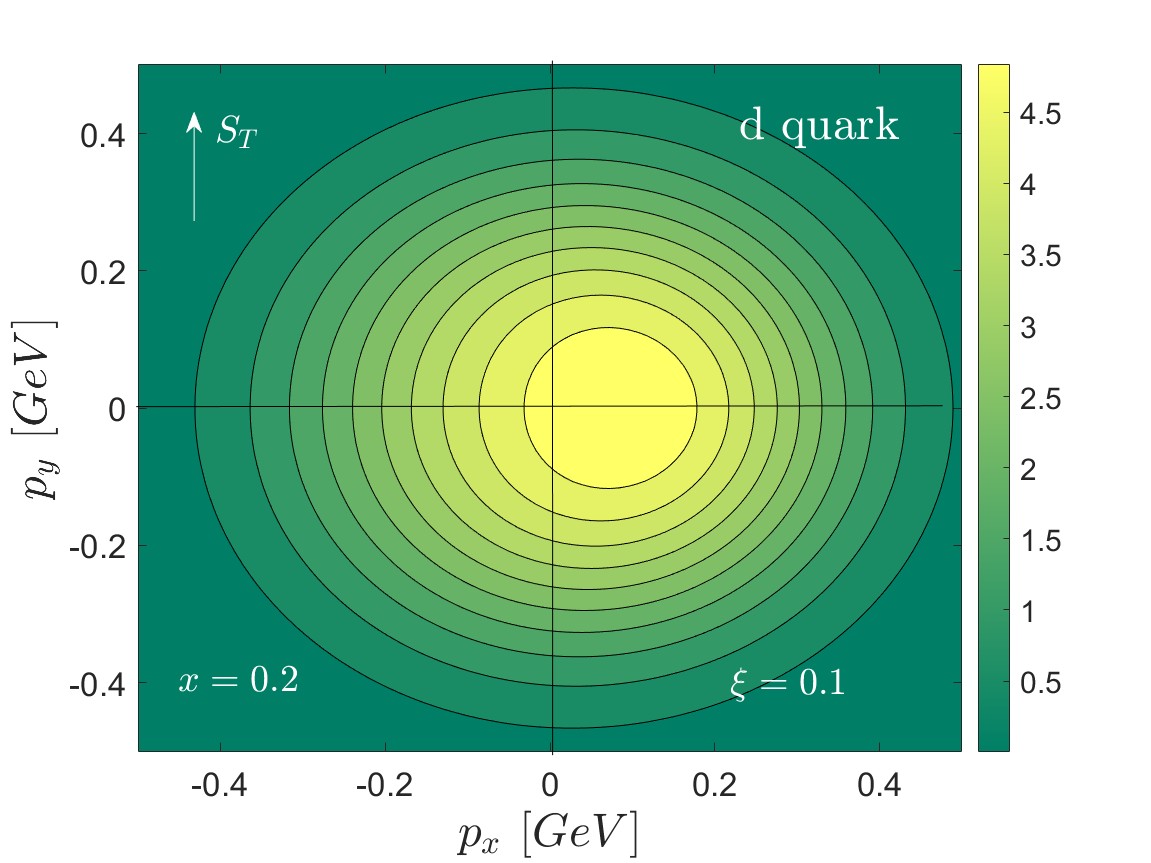}
\caption{Spin density for $u$-quark and $d$-quark in a transversely polarized proton with  at $x=0.2, \xi=0.1$ and for the limit $\Dp = 0$. The transverse spin $S_\perp$ is taken along $\hat{y}$.}\label{fig_spinD}
\end{figure}

The spin density of an unpolarised quark of skewness $\xi$ in a transversely polarised proton can be defined in terms of GTMDs at the vanishing transverse momentum transfer limit $\bfd=0$ as 
\be
f_{\nu/P^\uparrow}(x, \xi, \bfp) = \frac{1}{\zf} F^e_{1,1}(x,\xi, 0, \bfp^2, 0) - \frac{{\bf S}.(\hat{{\bf P}}\times\bfp)}{M} \zf F^{o}_{1,2}(x, \xi, 0,\bfp^2, 0), 
\ee
which satisfies the spin density defined in Ref.\cite{Bacchetta:2004jz} at $\xi=0$ limit. In the convention of Ref.\cite{Anselmino:2002pd}, the Siver effect is described with a sign difference of the distribution.  The non-vanishing T-odd GTMD $F^{o}_{1,2}$ provides correlation of proton spin to transverse momentum of the parton. We are interested in the effects of skewness $\xi$ on the spin-momentum correlation and the spin density. The model results for spin density are shown in Fig.(\ref{fig_spinD}), where the momentum of target proton $\hat{P}$ is considered along the momentum transfer $z$-axis and proton spin $S$ is taken along $\hat{y}$. As per the expectation, the model results are not symmetric and show a left shift for $u$ and right shift for $d$ quarks. Such left-right asymmetry can be explained from the numerical results, shown in Fig.\ref{fig_GTMD_xz}, for T-odd GTMDs $ F^{o}_{1,2}$ of Eq.(\ref{F12}).
The mean-square transverse momentum of the unpolarized quarks in a transversely polarized proton associated with spin density is given by
\be 
\langle \bfp^2 \rangle(\xi) = \int dx ~\int d^2\bfp~ \bfp^2  ~ f_{\nu/P^\uparrow}(x, \xi, \bfp) 
\ee
which is sensitive to the skewness variable $\xi$. 
 In this model, $\langle \bfp^2 \rangle^u=\{0.131, 0.108, 0.081\}$ and $\langle \bfp^2 \rangle^d=\{0.075, 0.060, 0.042\}$ corresponding to the skewness $\xi=\{0,0.1,0.2\}$ respectively. Similarly, the model results for the spin density of transversely polarized quarks in an unpolarized proton show an unidirectional left-shift for both $u$ and $d$ quarks as expected from the polarity non-flip distribution $H^{(0)}_{1,1}$, Fig.\ref{fig_GTMD_xz}.


\section{Sivers and Boer-Mulders WDs in boost invariant longitudinal position space \label{sec_WD_sig}}
 
 In this section, we discuss T-odd Wigner Distributions in boost-invariant longitudinal position space $\sigma$. 
The six-dimensional position-momentum distributions are found from the Fourier transformation of unintegrated quark-quark correlator $W^{\nu [\Gamma]}_{[\lambda^{\prime\prime}\lambda^{\prime}]}(x,\xi,\bfd^2, \bfp^2, \bfp.\bfd)$ where, transverse momentum transfer $\bfd$ is the Fourier conjugate to impact parameter space $\bfD=\frac{\bfb}{(1-\xi^2)}$. The boost-invariant longitudinal position space $\sigma$ is the Fourier conjugate to longitudinal momentum transfer $\xi$ and defined as $\sigma= \frac{1}{2}b^- P^+$. In case of quark-spectetor system, $\sigma$ is the dimensionless frame-independent version of the difference in longitudinal positions between the quark and spectator \cite{Miller:2019ysh}. The longitudinal boost invariant space is presented in a different notation $\Tilde{z}$ (with a 1/2 factor difference) in Refs.\cite{Miller:2019ysh,Weller:2021wog,Yang:2025neu}.
The Wigner distributions in boost-invariant position space are defined as \cite{Brodsky:2006in,Maji:2022tog}
\be
\tilde{\rho}^{\nu [\Gamma]}(x,\sigma,\bfd,\bfp;S)=\int^{\xi_{s}}_0 \frac{{\rm d} \xi}{2\pi}\, e^{i\sigma.\xi} \,W^{\nu [\Gamma]}(x,\xi,\bfd^2,\bfp^2,\bfd.\bfp;S)\,,
\label{wig_rho_sig}
\ee
where, the Wigner correlator $W^{\nu [\Gamma]}(x,\xi,\bfd^2,\bfp^2,\bfd.\bfp;S)$ is given in Eq.(\ref{Wdef}) with $\lambda^{\prime\prime}=\lambda^\prime= S$. 
 In this Fourier transformation, finiteness of the upper-limit $\xi_s$ is restricted by the allowed range given by $ \xi_{max}$ as
\be 
\xi_{max} = \frac{(-t)}{2M^2} \left(\sqrt{1+\frac{4M^2}{(-t)}} -1\right)\label{zmax},
\ee                                                                                                                                                                                                                                                                                                                                      
where $M$ is the mass of target proton \cite{Brodsky:2000xy, Chakrabarti:2008mw,Manohar:2010zm}.
We restrict ourself to the DGLAP region $\xi<x<1$ and the limit of the Fourier transform in $\sigma$-space is taken as  $\xi_s=\xi_{max}$ if $\xi_{max}<x$ and  $\xi_s=x$ for $\xi_{max}>x$, determined by a fixed value of $-t$.
Following the similar prescription for polarization configurations of Wigner distributions in the $\bfb$ space ~\cite{Lorce:2011kd,Liu:2015eqa,Chakrabarti:2017teq}, the WDs in the longitudinal position space are defined as
\be 
\tilde{\rho}^{i \nu }_{TU}(x,\sigma,\bfd,\bfp)&=&\frac{1}{2}[\tilde{\rho}^{\nu [\gamma^+]}(x,\xi,\bfd^2,\bfp^2,\bfd.\bfp; +\hat{S}_i) - \tilde{\rho}^{\nu [\gamma^+]}(x,\xi,\bfd^2,\bfp^2,\bfd.\bfp; -\hat{S}_i)]\,,\label{rho_TU_def}\\
\tilde{\rho}^{j \nu }_{UT}(x,\sigma,\bfd,\bfp)&=&\frac{1}{2}[\tilde{\rho}^{\nu [i\sigma^{j+}\gamma^5]}(x,\xi,\bfd^2,\bfp^2,\bfd.\bfp; +\hat{S}) - \tilde{\rho}^{\nu [i\sigma^{j+}\gamma^5]}(x,\xi,\bfd^2,\bfp^2,\bfd.\bfp; -\hat{S})]\,,\label{rho_UT_def}
\ee
where, transverse index $i,j=\{1,2\}$ for $x$ and $y$ direction respectively, and the transverse spin of the proton is defined as $\hat{S}_j=\frac{1}{\sqrt{2}}\left(|+S \rangle + |-S\rangle  \right) $. The subscripts of $\tilde{\rho}^{j \nu }_{XY}$ represent polarization of proton (by $X$) and of interior quark (by $Y$) respectively e.g., $\tilde{\rho}^{j \nu }_{TU}$ represents WDs of unpolarized quarks in a transversely polarised proton and $\tilde{\rho}^{j \nu }_{UT}$ is for the transversely polarised quark in unpolarised proton. For $\tilde{\rho}^{j \nu }_{TU}$, only the off-diagonal terms of Correlator $W^{\nu [\Gamma]}_{[\lambda^{\prime\prime}~\lambda^\prime]}$ contribute while for $\tilde{\rho}^{j \nu }_{UT}$ the diagonal terms contribute and can be written as
\be
\hat{\rho}^{i\nu}_{TU}(x, \sigma, t, \bfp)&=& - \frac{1}{2} \int^{\xi_{s}}_0 \frac{{\rm d} \xi}{2\pi}\, e^{i\sigma.\xi} ~ \epsilon^{i j}_\perp (i)^j \left[ W^{\nu [\gamma^+]}_{[+ -]}(x,\xi,\bfd^2, \bfp^2, \bfp.\bfd)+ (-1)^j~ W^{\nu [\gamma^+]}_{[- +]}(x,\xi,\bfd^2, \bfp^2, \bfp.\bfd)\right] \nonumber \\
&=& \tilde{\rho}^{i\nu}_{TU}(x, \sigma, t, \bfp)  +\frac{1}{M}\epsilon^{ij}_\perp p^j_\perp  \tilde{\rho}^\nu_{Siv}(x, \sigma, t, \bfp), \label{RUTSiv}\\
\hat{\rho}^{j \nu}_{UT}(x, \sigma, t, \bfp)&=& \frac{1}{2} \int^{\xi_{s}}_0 \frac{{\rm d} \xi}{2\pi}\, e^{i\sigma.\xi} ~ \left[ W^{\nu [i\sigma^{j+}\gamma^5]}_{[+ +]}(x,\xi,\bfd^2, \bfp^2, \bfp.\bfd)+ W^{\nu [i\sigma^{j+}\gamma^5]}_{[- -]}(x,\xi,\bfd^2, \bfp^2, \bfp.\bfd)\right] \nonumber \\
&=& \tilde{\rho}^{j \nu}_{UT}(x, \sigma, t, \bfp) -\frac{1}{M}\epsilon^{ij}_\perp p^i_\perp \tilde{\rho}^\nu_{BM}(x, \sigma, t, \bfp). \label{RUTBM}
\ee
The T-odd coefficients of transverse quark momentum $\bfp$ terms of the above Wigner Distributions in longitudinal boost invariant space are separated as Sivers and Boer-Mulder and defined as 
\be
\tilde{\rho}^\nu_{Siv}(x, \sigma, t, \bfp) &=&- \int_0^{\xi_s}\frac{d\xi}{2\pi} e^{i\sigma.\xi}~ \zf  ~F^{o \nu}_{1,2}(x, \xi, t, \bfp^2, \bfd.\bfp),\label{SiSiv}\\
\tilde{\rho}^\nu_{BM}(x, \sigma, t, \bfp) &=& -\int_0^{\xi_s} \frac{d\xi}{2\pi} e^{i\sigma.\xi} ~  \frac{1}{\zf}~ H^{o \nu}_{1,1}(x, \xi, t, \bfp^2, \bfd.\bfp), \label{SiBM}
\ee
These play a crucial role in the correlation between spin and transverse momentum that led to the single-spin asymmetries. The above definition of the Sivers and Boer-Mulder distribution provide pictures in boost invariant longitudinal position space of the T-odd sector, which are useful for the experimental measurements. The real-life experiments are restricted to non-zero skewness measurements of the distributions.  
  The transverse energy transfer $\Dp^2$ is replaced by the total momentum transfer $t$ using the relation Eq.(\ref{mt_def}). In the above Eqs.(\ref{SiSiv}, \ref{SiBM}), the minus sign is included to maitain the consistency with the definition in TMD limit.  
The complete expression of the left-out part of the WDs $\tilde{\rho}^{i\nu}_{TU}(x, \sigma, t, \bfp)$ and $\tilde{\rho}^{j\nu}_{UT}(x, \sigma, t, \bfp)$ are included in App.\ref{AppA}.

\begin{figure}[h]
\includegraphics[scale=.35]{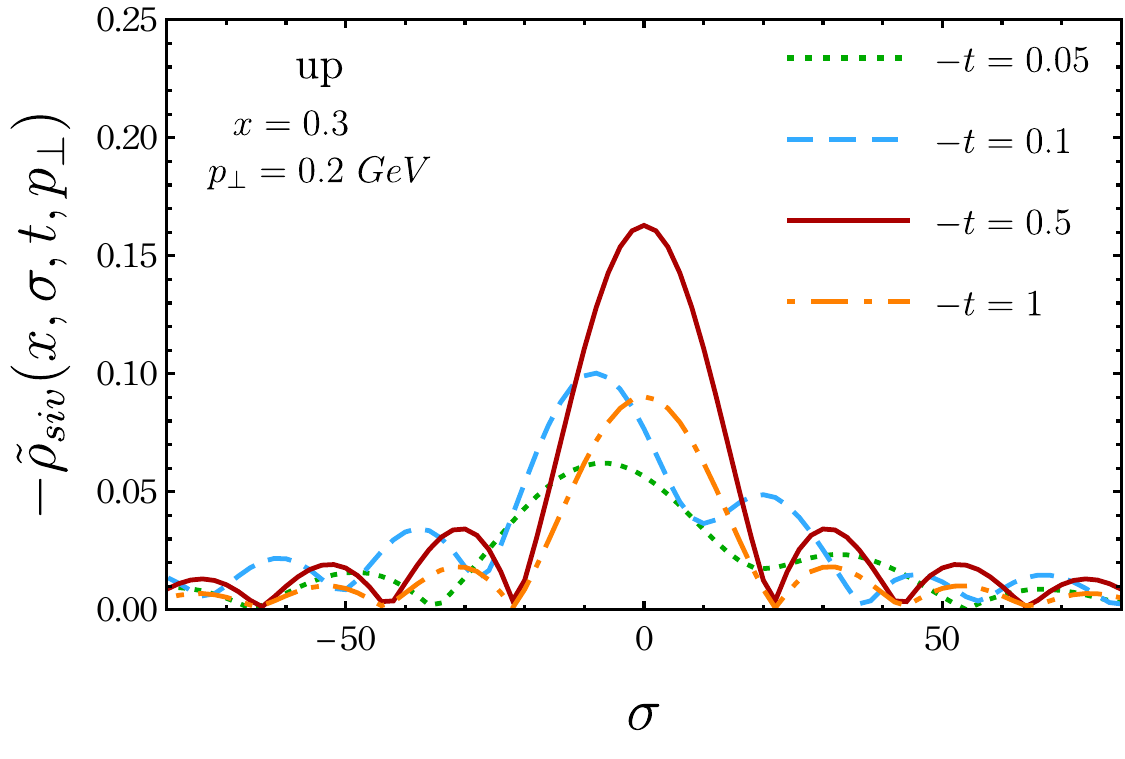} 
\includegraphics[scale=.35]{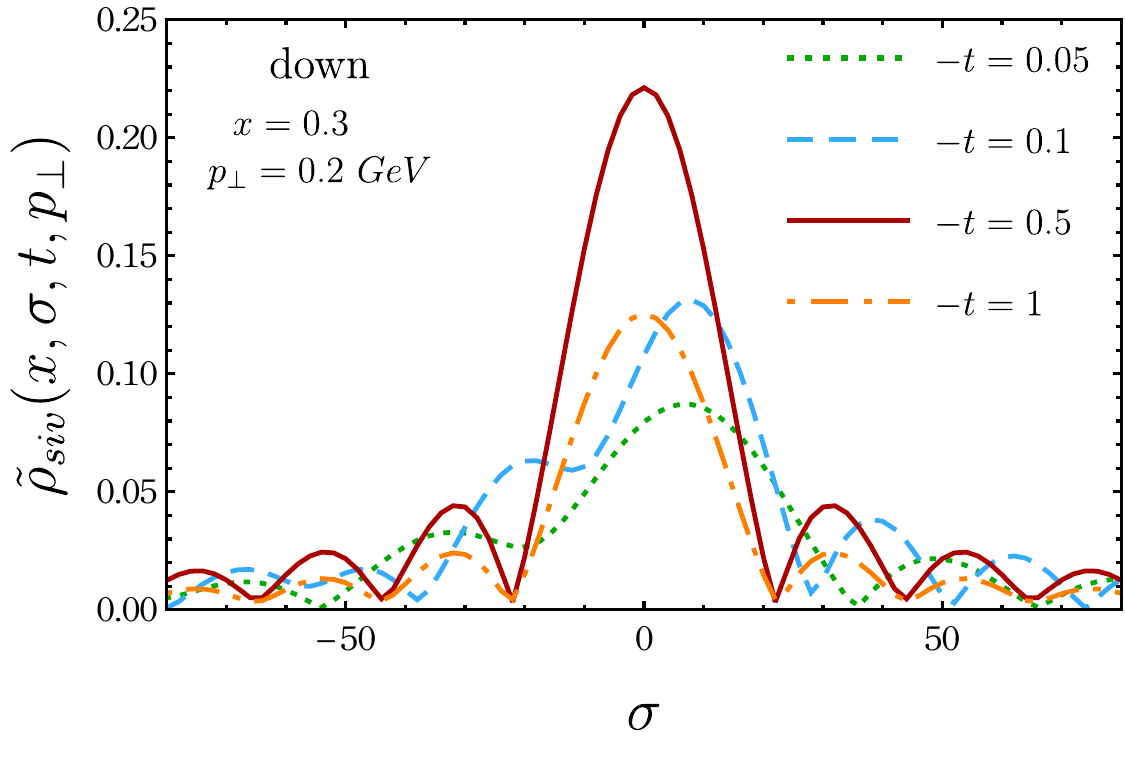}
\includegraphics[scale=.35]{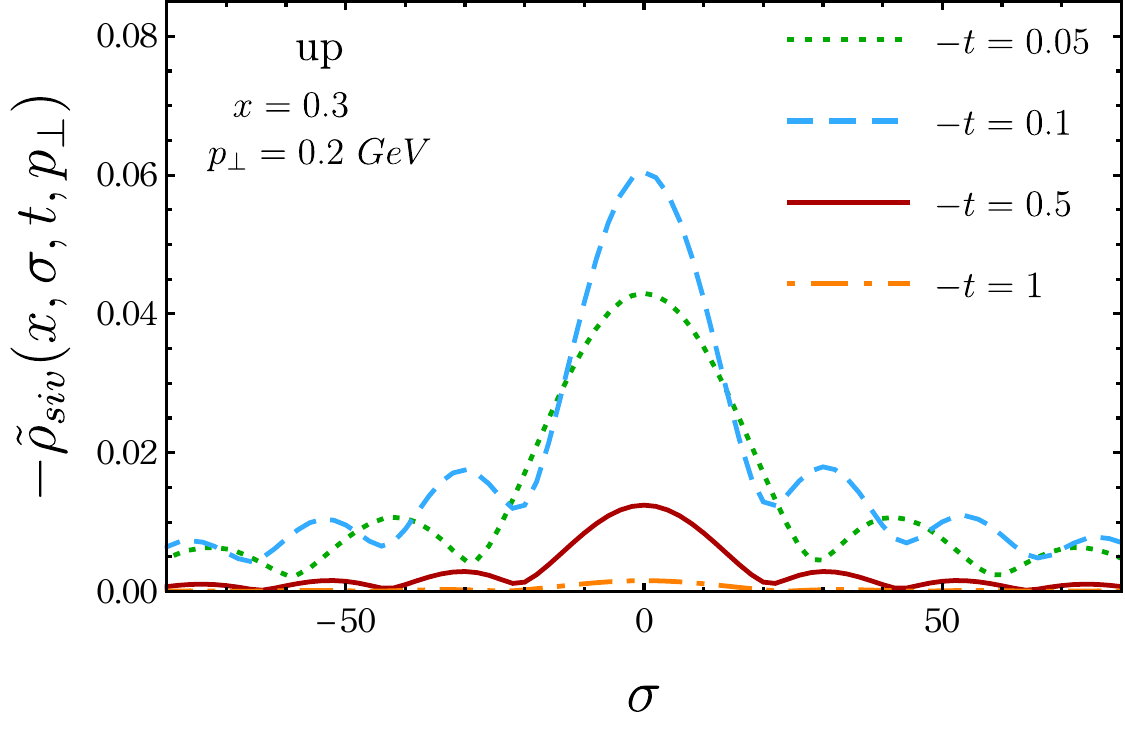} 
\includegraphics[scale=.35]{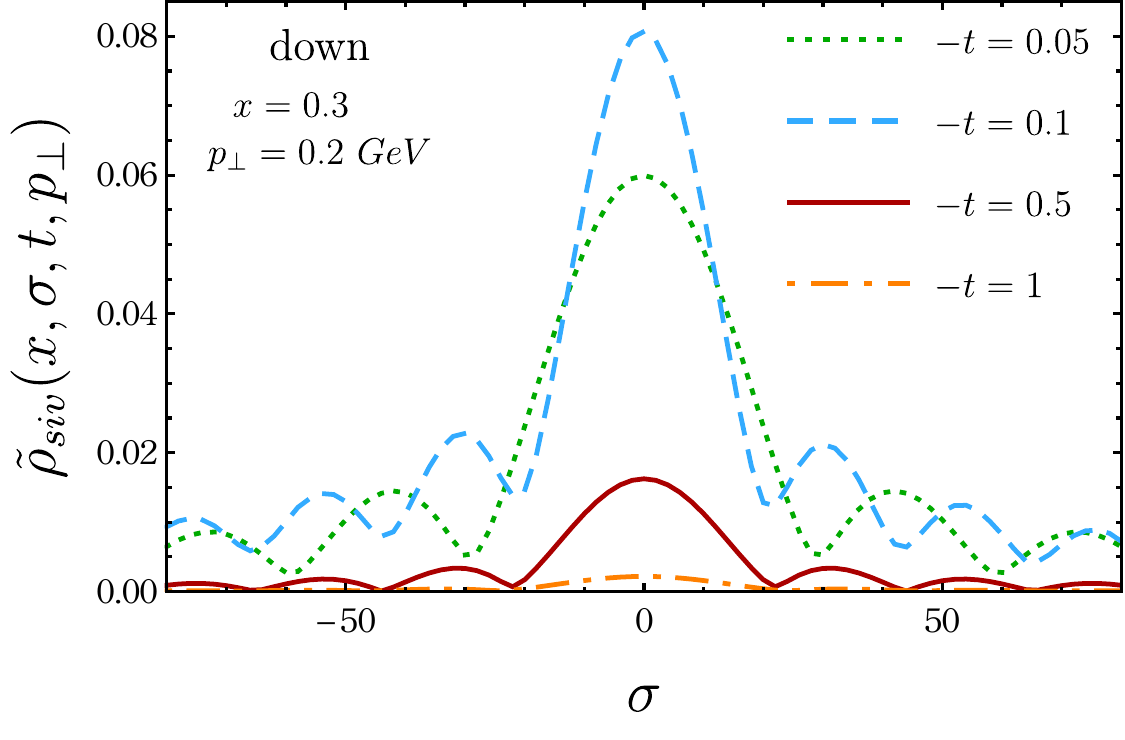}
\caption{Sivers Wigner distribution in boost invariant longitudinal space at $x=0.3, \bfp=0.2\hat{p}_y~GeV$ along $\hat{y}$. The four plots in each sub-figure are for $-t= 0.05, 0.1, 0.5, ~and ~ 1$. The left and right columns are for $u$-quark and $d$-quark, respectively. The upper row shows the model results when $\bfp$ and $\bfd$ are parallel, whereas the lower row is when $\bfp$ and $\bfd$ are perpendicular, i.e., $\bfp.\bfd=0$. }\label{fig_Siv_sig}
\end{figure}

In Fig.\ref{fig_Siv_sig}, the numerical plots of Sivers distribution $\tilde{\rho}^\nu_{Siv}(x, \sigma, t, \bfp)$ is shown in the boost invariant longitudinal at $x=0.3, \bfp=0.2 \hat{p}_y ~GeV$ space for $u$ (left) and $d$ (right) quarks. Note that, the Sivers WDs in $\sigma$-space are negative for $u$ quark, and a minus sign is included in the axis label. In each of the plots, four different color-codes(line codes) are corresponding to the four fixed values of $-t= 0.05,0.1, 0.5~and~ 1$. 

The distribution in boost invariant longitudinal space shows oscillating behaviour, which is analogous to the optical diffraction phenomenon. The amplitude of the maxima of Sivers WDs in $\sigma$-space is larger for $d$-quark than for the $u$-quark. Finiteness of the width of a slit is a necessary condition to achieve a diffraction pattern in Optics. Here, the finite limit of the Fourier integration in $\xi_s$ (from $0$ to $\xi_{max}$ or $x$) plays an equivalent role to the slit width for the occurrence of the diffraction pattern. For all the plots in $\sigma$-space, the width of principle maxima becomes narrower with the increase of $\xi_{max}$ determined by larger energy transfer $-t$. This behaviour is similar to optical diffraction through slit where the position of the first minima varies inversely to the slit width.

For last two values of $-t=\{0.5,1.0\}$, from Eq.(\ref{zmax}) the corresponding  $\xi_{max}=\{0.521, 0.639\}$ respectively, which belong to the region $\xi_{max}>x$ and the limit of the Fourier transform in Eq.(\ref{SiSiv}) set to $\xi_s=x=0.3$ for both the cases. As the upper limit of the Fourier transformation is fixed to $x=0.3$, no changes are noticed in the position of the first minima for $-t=\{0.5,1.0\}$ marked by red continuous and orange dot-dashed lines. However the peak value of the central maxima changes significantly, and the position of the central maxima reflects a symmetric pattern for both the quarks. The slit width causing the diffraction pattern in $\sigma$-space is determined dynamically by the kinematics $-t$ and the longitudinal momentum fraction $x$.

\begin{figure}[t]
\includegraphics[scale=.35]{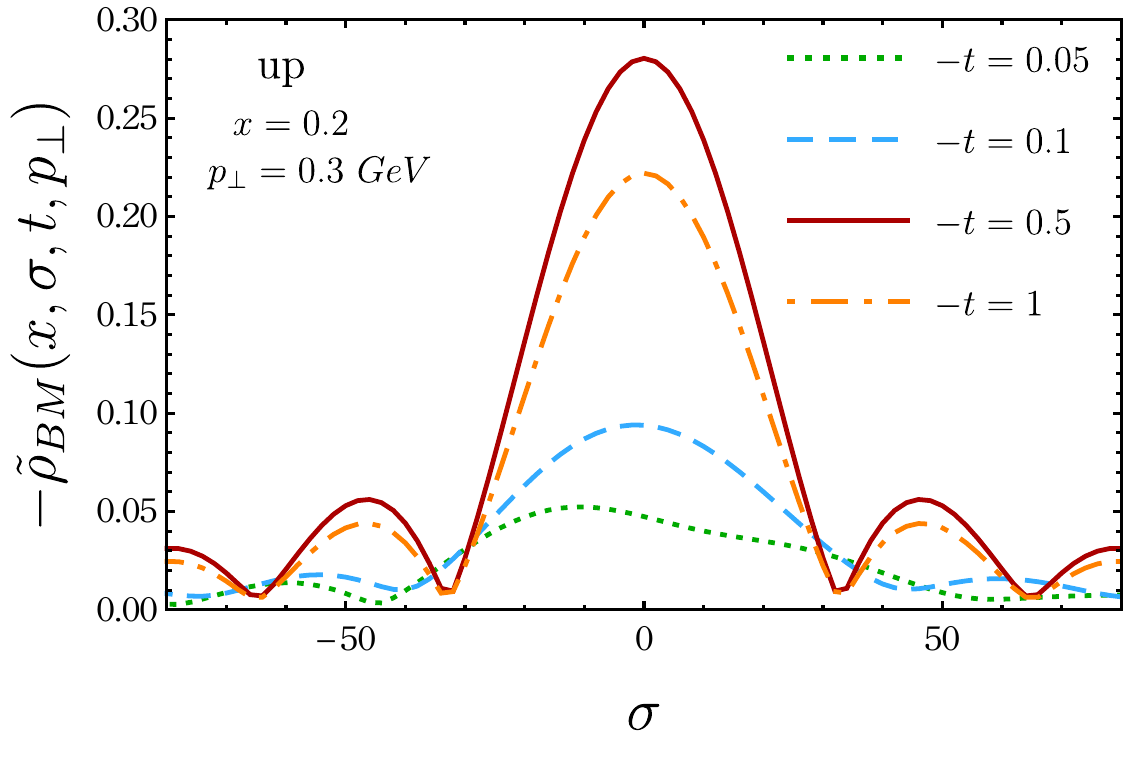} 
\includegraphics[scale=.35]{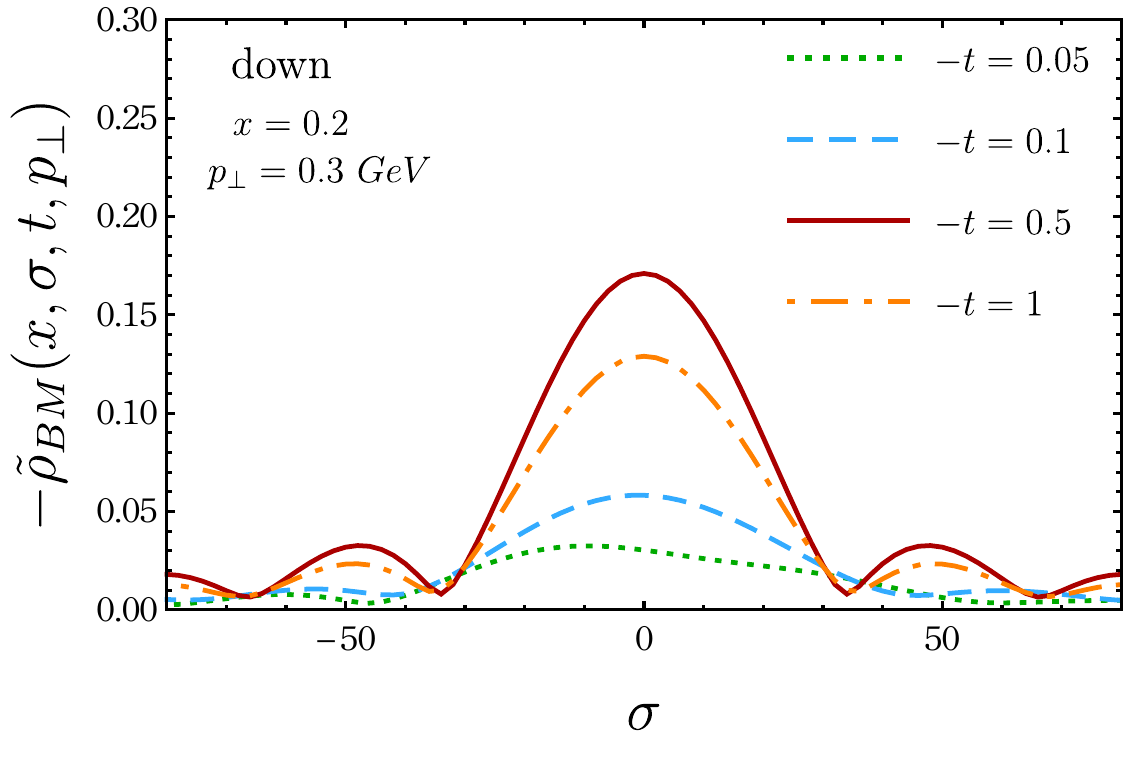}
\includegraphics[scale=.35]{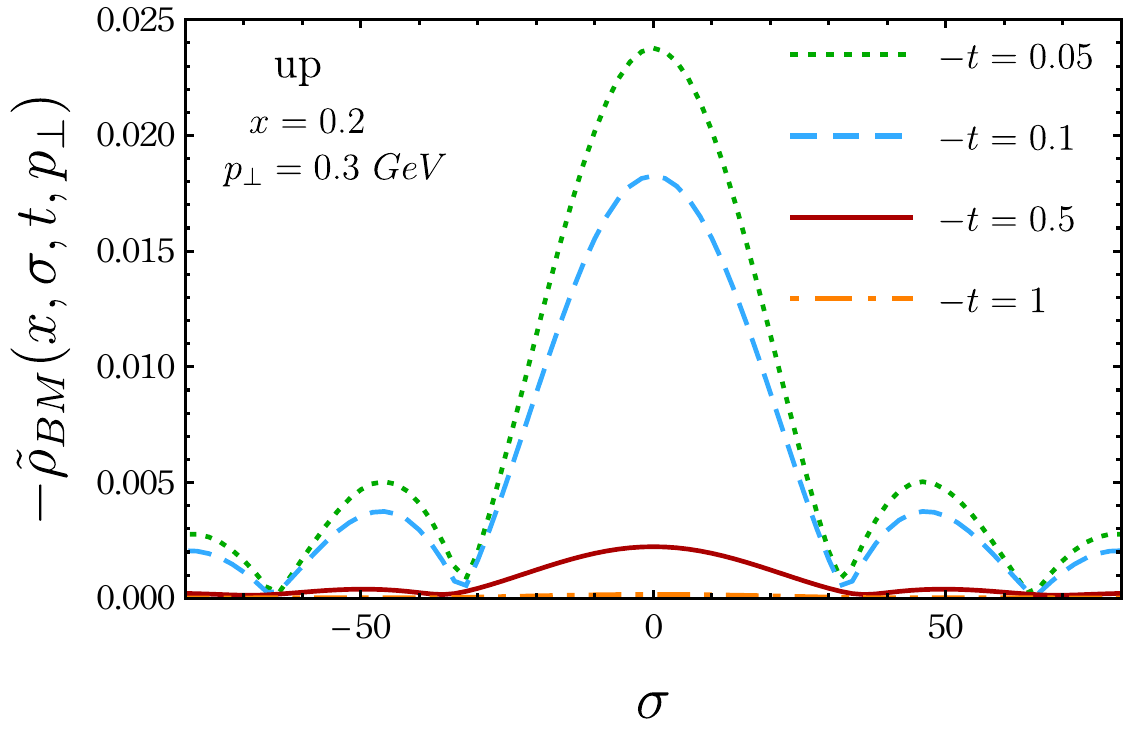} 
\includegraphics[scale=.35]{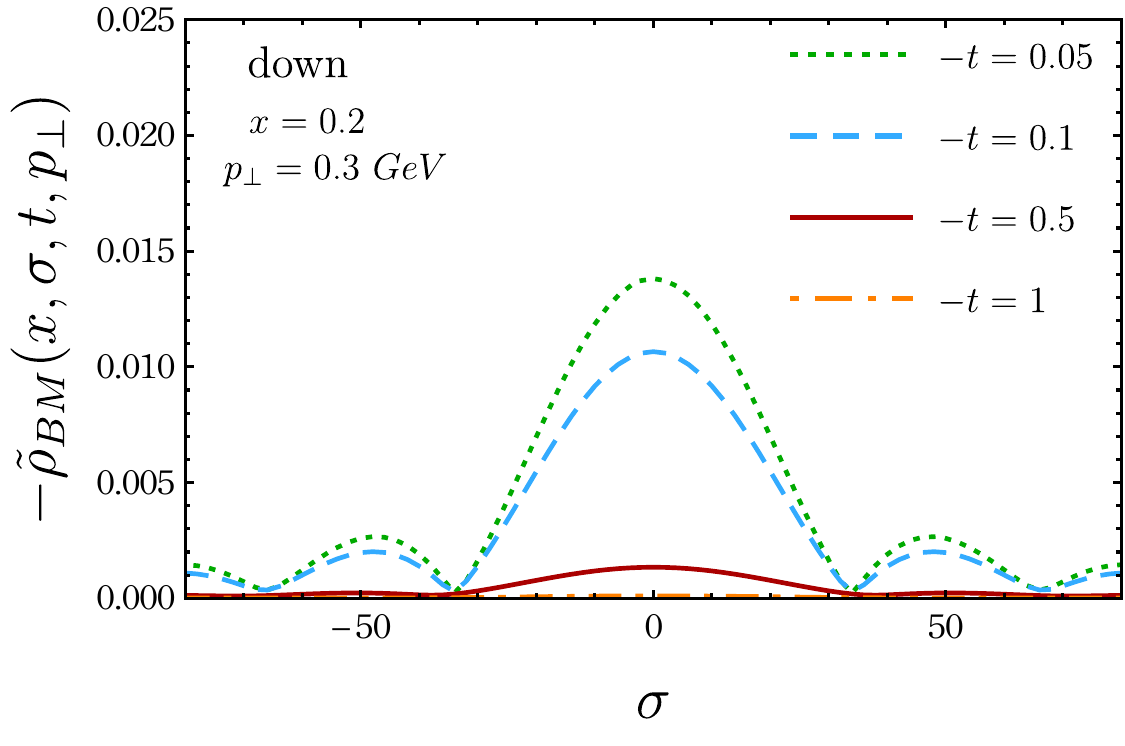}
\caption{Boer-Mulders Wigner distribution in boost invariant longitudinal space at $x=0.3, \bfp=0.2\hat{p}_y~GeV$ along $\hat{y}$. The four plots in each sub-figure are for $-t= 0.05, 0.1, 0.5, ~and ~ 1$. The left and right columns are for $u$-quark and $d$-quark, respectively. The upper row shows the model results when $\bfp$ and $\bfd$ are parallel, whereas the lower row is when $\bfp$ and $\bfd$ are perpendicular, i.e., $\bfp.\bfd=0$. }\label{fig_BM_sig}
\end{figure} 

For the first two choices of $-t= \{0.05, 0.1\}$, $\xi_{max}=\{0.211, 0.285\}$ respectively, which 
lie in the region $\xi_{max}<x$ (as $x=0.3$) and the integration limit is set to $\xi_s=\xi_{max}$. Thus, the position of the first minima for $-t=0.1$ is smaller than the position of the first minima for $-t=0.05$. In other word, the width decreases as $-t$ increases in this region. Additionally, we have noticed that the pattern is no longer symmetric and show left shifting for $u$ and right shifting for $d$ quarks, which are consistent with the left-right shift in quark density associated to the Sivers effect, as shown in Fig.\ref{fig_spinD}. We also have noticed a significant secondary maxima for both the quarks, which is due to the dominating interference among $\bfp$ and $\bfd$ in this region $\xi_{max}<x$. Here, the upper row of Fig.\ref{fig_Siv_sig} shows our model results when $\bfp$ is chosen parallel to $\bfd$. 
The terms contain $\bfp.\bfd$ led to an asymmetric diffraction pattern, which can be interpreted as an additional significant contribution from the interference over diffraction as found in Optics. For completeness, the Sivers distribution in $\sigma$-space is investigated by vanishing the interference term by putting $\bfp.\bfd=0$ in Eq.(\ref{SiSiv}) and shown in Fig.\ref{fig_Siv_sig}(lower row). The $\bfp.\bfd$ term generates from the square of transverse momentums of incoming  $\bfp^\prime$ (Eq.\ref{ptp}) and out going $\bfp^{\prime\prime}$ (Eq.\ref{ptpp})  proton. Eventually, all the $(\chi_i-\chi_j)$ terms and the exponential factor contain $\bfp.\bfd$ term. We have noticed that, for the region $\xi_{max}<x$, the distribution becomes symmetric and the width of the central maxima decreases as the total momentum transfer $-t$ increases. The overall amplitude reduced by a large amount which reflects a significant amount of contribution is gettered from the $\bfp.\bfd$ term in boost invariant longitudinal space.  
Thus, the diffraction pattern is not solely due to the finite limit of Fourier $\xi$ integration, and the functional forms of the GTMDs are also important for this phenomenon. 

This kind of oscillating behavior is not very exceptional; a similar paraxial optics and quantum fields on the light cone were first reported in \cite{Sudarshan:1982gv, Mukunda:1982gu}. Notably, distributions in the boost invariant longitudinal space show a long-distance tail as reported recently in Refs.\cite{Miller:2019ysh,Weller:2021wog}. 
A qualitatively similar diffraction pattern has also been observed in other observables such as DVCS amplitude, GPDs, and the parton density in longitudinal position space  \cite{Brodsky:2006in,Brodsky:2006ku,Chakrabarti:2008mw,Manohar:2010zm}.

Fig.\ref{fig_BM_sig}, shows our model result of the Boer-Mulder Wigner distribution in boost invariant longitudinal position space for $u$ and $d$ quarks. A similar diffraction pattern is noticed, caused by the finite integration limit in the Fourier transformation in Eq.(\ref{SiBM}). 
The model result for the Boer-Mulders Wigner distribution is negative for both the quarks, while the peaks of maxima are larger for $u$ quark than the $d$ quark.  The change in polarity from $u$ to $d$ quark justifies the left-right symmetry for the Sivers function, whereas Boer-Mulders distribution does not show any change in polarity which led to a one-directional shift in spin density. This difference can be justified from the model results of $F_{1,2}$ and $H_{1,1}$ GTMDs shown in Fig.\ref{fig_GTMD_xz}. The upper row is the model result when $\bfp$ is parallel to $\bfd$, and a less significant interference is noticed for the region $\xi_{max}<x$ compared to the Sivers distribution. This can be understood from the model result of $H_{1,1}$ which does not have the third term (Eq.\ref{H11}) containing $(\chi^{\prime\prime}_2-\chi^{\prime}_2)$ term like Sivers distribution. The lower row of Fig.\ref{fig_BM_sig} represents the model results when $\bfp$ and $\bfd$ are perpendicular to each other and the distribution becomes symmetric as expected.

%
%

\section{T-odd Wigner distributions in $\bfb$-space \label{sec_WD_bt}}
For completeness, in this section, we present the T-odd WDs in transverse impact parameter space $\bfb$. The impact parameter $\bfb$ is conjugate to $\bfD= \bfd/(1-\xi^2)$ and using this relation the $\bfd$ is replaced by $\bfD$.  
The real part of the modified Wigner distribution  $\bar{\rho}^{i\nu}_{TU}(x, \xi, \bfb, \bfp)$ is given by
\be
\bar{\rho}^{i\nu}_{TU}(x, \xi, \bfb,\bfp)&=& - \frac{1}{2} \int \frac{d^2\bfD}{(2\pi)^2} e^{-i\bfD. {\bfb}} \epsilon^{i j}_\perp (i)^j \left[ W^{\nu [\gamma^+]}_{[+ -]}(x, \xi, \bfD,\bfp)+ (-1)^j~ W^{\nu [\gamma^+]}_{[- +]}(x, \xi, \bfD,\bfp)\right] \nonumber \\
&=& \rho^{i\nu}_{TU}(x, \xi, \bfb,\bfp)  +\frac{1}{M}\epsilon^{ij}_\perp p^j_\perp  \bar{\rho}^\nu_{Siv}(x, \xi, \bfb, \bfp), \label{RUTSiv_bt}\\
\bar{\rho}^{j \nu}_{UT}(x, \xi, \bfb,\bfp)&=& \frac{1}{2} \int \frac{d^2\bfD}{(2\pi)^2} e^{-i\bfD. {\bfb}} \left[ W^{\nu [i\sigma^{j+}\gamma^5]}_{[+ +]}(x, \xi, \bfD,\bfp)+ W^{\nu [i\sigma^{j+}\gamma^5]}_{[- -]}(x, \xi, \bfD,\bfp)\right] \nonumber \\
&=& \rho^{j \nu}_{UT}(x, \xi, \bfb,\bfp) -\frac{1}{M}\epsilon^{ij}_\perp p^i_\perp \bar{\rho}^\nu_{BM}(x, \xi, \bfb, \bfp). \label{RUTBM_bt}
\ee
Where, $\rho^{i\nu}_{TU}(x, \xi, \bfb,\bfp) $ and $\rho^{j \nu}_{UT}(x, \xi, \bfb,\bfp)$ are the T-even part without the FSI contribution and include the explicit forms in Appendix-\ref{AppA}. 
The second terms in Eqs.(\ref{RUTSiv_bt},\ref{RUTBM_bt}) are T-odd contributions that can be considered as the phase space correlation map to the Sivers and Boer-Mulders Wigner distributions (Eq.(\ref{RUTSiv},\ref{RUTBM})) respectively as
\be
\bar{\rho}^\nu_{Siv}(x, \xi, \bfb, \bfp) &=&- \int\frac{d^2\bfD}{(2\pi)^2} e^{-i\bfD.\bfb} \zf  ~F^{o \nu}_{1,2}(x, \xi, \bfD, \bfp),\label{RSiv_bt}\\
\bar{\rho}^\nu_{BM}(x, \xi, \bfb, \bfp) &=& -\int\frac{d^2\bfD}{(2\pi)^2} e^{-i\bfD.\bfb}   \frac{1}{\zf}~ H^{o \nu}_{1,1}(x, \xi, \bfD, \bfp), \label{RBM_bt}
\ee 
The Sivers effect refers to a shift in momentum space ($k_\perp$) describes a correlation between transverse spin of a proton and the transverse momentum of unpolarized constituent partons. 

\begin{figure}[htb] 
\hspace{-1cm}
\includegraphics[scale=.3]{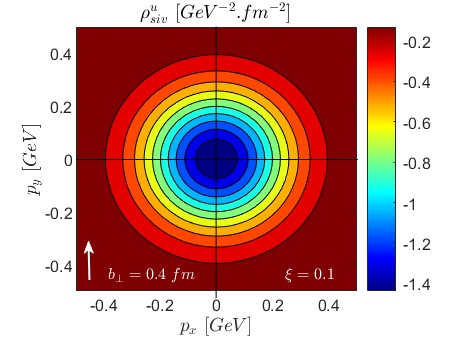}\hspace{-.3cm}
\includegraphics[scale=.3]{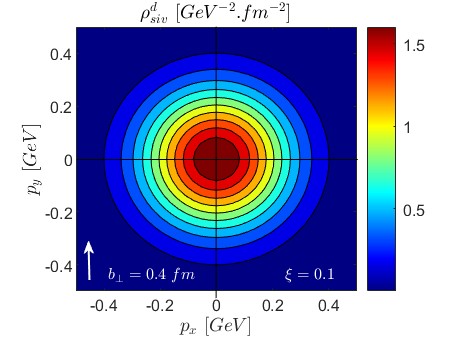}\hspace{-.3cm}
\includegraphics[scale=.3]{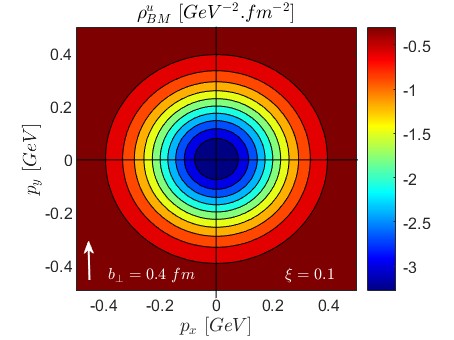} \hspace{-.3cm}
\includegraphics[scale=.3]{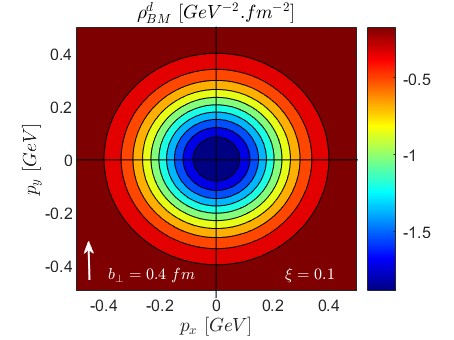}\\ \hspace{-1cm}
\includegraphics[scale=.3]{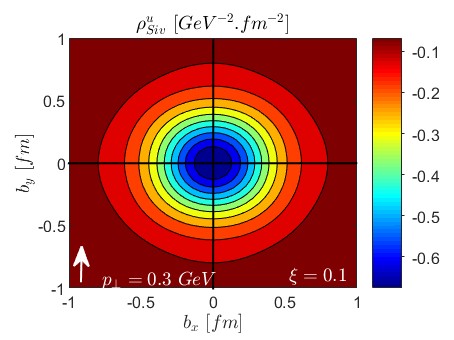}\hspace{-.3cm}
\includegraphics[scale=.3]{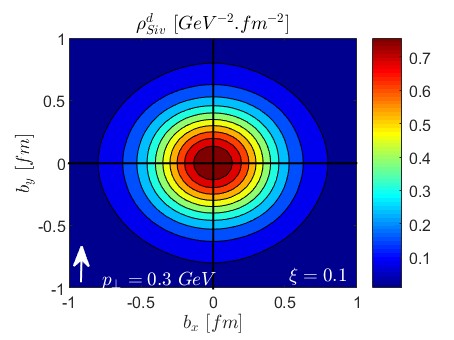} \hspace{-.3cm}
\includegraphics[scale=.3]{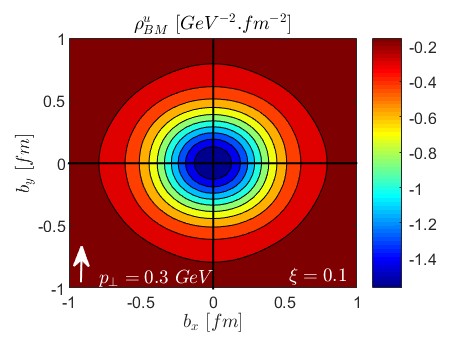} \hspace{-.3cm}
\includegraphics[scale=.3]{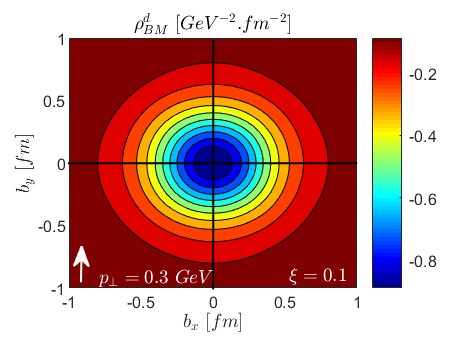}
\caption{The flavor-dependent Sivers and Boer-Mulders Wigner Distribution in transverse momentum plane (upper row) with $\bfb=0.4~\hat{y}$ and transverse impact parameter plane (lower row) with $\bfp = 0.3~GeV$ at skewness $\xi=0.1$.  }\label{fig_siv_btpt}
\end{figure} 

We investigate these functions numerically in the transverse momentum and transverse position space. The model results for Sivers function as a function of transverse impact parameter space are shown in Fig.\ref{fig_siv_btpt}. The upper row represents momentum space distribution with non-zero skewness $\xi=0.1$ at $\bfb=0.4~fm$ along $\hat{y}$ for both the flavors. While the lower row is for transverse impact parameter plane with skewness $\xi=0.1$ at $\bfp=0.3~GeV$ along $\hat{y}$. We notice polarity flip with the change in flavor in both the $\bfp$ and $\bfb$ planes, which are at per our expectation of model results of positive and negative  $F^o_{1,2}$ distributions under flavor change. 
For numerical calculation, we choose the transverse momentum transfer $\bfd$ is parallel to the transverse momentum to the quark $\bfp$. The circularly symmetric contour plots in $\bfp$-plane indicates insignificant contribution from the crossing terms containing $\bfp. \bfd$. While the transverse impact parameter plane $\bfb$-plane distribution is no longer circularly symmetric, caused by the terms containing $\bfp.\bfd$ and shows an axial symmetry.   
We also noticed that, for both the plane, intensity of the Sivers WDs decreases for increasing skewness $\xi$. The model results for the Boer-Mulders distribution show similar patterns-- $\bfp$-plane is circularly symmetric whereas $\bfb$-plane is axially symmetric, as shown in Fig.\ref{fig_siv_btpt}. Note that, Sivers distributions is part of the Wigner distribution for unpolarised quark in a transversely polarised proton $\bar{\rho}^{i\nu}_{TU}(x, \xi, \bfb,\bfp)$, which shows an dipole structure in $\bfb$-plane caused by the transverse momentum cross-terms present in first part of Eq.(\ref{RUTSiv_bt}). The axially symmetric Sivers distribution is considered as an enhancement in intensity, preserving the di-polar structure in $\bfb$-plane. The intensity of the distributions is sensitive to the skewness given by this model may be useful for phenomenological study of experimental measurements.





\section{Conclusions}\label{con}
Here, we present the skewness sensitivity of the T-odd GTMDs e.g., Sivers and Boer-Mulders Wigner distributions, considering the momentum transfer in both the transverse and the longitudinal directions. 
 The primary focus of this work is a detailed model study on Sivers and Boer-Mulders WDs in boost invariant longitudinal position space $\sigma$, which is conjugate to the skewness $\xi$.  
 The leading twist T-odd quark GTMDs  
Protons are investigated using an extended light-front quark-diquark model motivated by soft-wall AdS/QCD. The T-odd sectors are accessed incorporating final state interaction (FSI) and introducing a phase to the wavefunctions.  
In both $\xi$-$x$ plane and $x$-$\bfd$ plane, the model results for GTMDs $F^o_{1,2}$ show a sign flip under a change in flavor, while the polarity of $H^o_{1,1}$ remain the same irrespective of flavor. 
All the numerical results are shown in the DGLAP region, i.e., for $\xi<x<1$.  
The magnitude of both $F^o_{1,2}$ and  $H^o_{1,1}$ die out with the increase of $\xi$ and peak shift towards right upto $x<0.5$.   
The Fourier transformation of $F^o_{1,2}$ and  $H^o_{1,1}$ over skewness, leads to a boost invariant longitudinal position space ($\sigma$) configuration of Sivers and Boer-Mulders Wigner distributions. The spin density of an unpolarized quark of skewness $\xi$ in a transversely polarized proton shows a left-right axial shift for $u$ and $d$ quarks. The mean-squre transverse momentum $<\bfp^2>$ decreases as the skewness value increases.

The LFQDM model results of Sivers and Boer-Mulders Wigner distributions in $\sigma$ space, 
show an oscillatory behavior which is analogous to the diffraction phenomenon in wave optics. The width of the central diffraction maxima is sensitive to the square of momentum transfer $-t$. As $-t$ increases, the width of principal maxima becomes narrower. Such an inverse variation in diffraction resembles the relation between slit-width and wavelength in Wave Optics. Here, the square of momentum transfer $-t$ plays a similar role to the effective slit-width. 
However, such a diffraction pattern is not solely due to the finiteness of the Fourier integration limit over $\xi$, but the functional behaviors of the GTMDs are crucial. In this model the correlation among $\bfp$ and $\bfd$ plays a crucial role-- if $\bfp$ is perpendicular to $\bfd$, the diffraction pattern is symmetric like a single-slit diffraction pattern, while if both are parallel to each other, the diffraction pattern loses its symmetry, Which can be thought of as an effect caused by interference between $\bfp$ and $\bfd$. Such an interference pattern demands other model calculations or future experimental measurements for verification. 
A similar diffraction pattern has also been observed in several other observables, such as DVCS amplitude, GPDs, and the parton density in longitudinal position space.

For completeness, the Sivers and Boer-Mulders WDs are also presented in transverse momentum space as well as in transverse impact parameter space. The model result shows a circular symmetry in $\bfp$-plane, whereas axial symmetry in $\bfb$-plane, which is considered as an additional contribution from T-odd sector to the complete Wigner Distribution of unpolarized quarks in a transversely polarized proton.

\begin{acknowledgments}
This work is supported
by the Anusandhan National Research Foundation (ANRF), Department of Science and Technology, Government of India and Science and Engineering Research Board (SERB) through the SRG (Start-up Research Grant) of File No.
SRG/2023/001093.

\end{acknowledgments}

\appendix

\section{}\label{AppA}

The spinors $u(k,\lambda) $ with the momentum $k$ and the helicity $\lambda\,(=\pm )$ are given by
\begin{center}
$u(k, +)=\frac{1}{\sqrt{2 k^+}}\left( \begin{matrix} 
  k^+ + m_F\\
  k^1 + i k^2\\
  k^+ - m_F\\
  k^1 + i k^2\\
\end{matrix} \right) $, \hspace{1cm}
$u(k, -)=\frac{1}{\sqrt{2 k^+}}\left( \begin{matrix} 
    - k^1 + i k^2\\
    k^+ + m_F\\
  k^1- i k^2\\
  - k^+ + m_F\\
\end{matrix} \right) $
\end{center}
with $m_F$ being the mass of the fermion.
Using the kinematics given in  Eqs.(\ref{Pp}), (\ref{Ppp}), one can find out the spinors $u(P^\prime,\lambda^\prime) $  and $u(P^{\prime \prime},\lambda^{\prime \prime}) $ and compute the matrix elements of $\bar{u}(k,\lambda)\Gamma u(k,\lambda)$, where $\Gamma$ represents the Dirac matrix structure.

The explicit expression of $\hat{\rho}^{j \nu}_{UT}(x, \sigma, t, \bfp)$ and $\hat{\rho}^{j \nu}_{TU}(x, \sigma, t, \bfp)$   in Eq.(\ref{RUTSiv}) and Eq.(\ref{RUTBM}) can be expressed in terms of GTMDs as  
\be
\tilde{\rho}^{i\nu}_{TU}(x,\sigma,t,\bfp)&=&  \int^{\xi_{s}}_0 \frac{d \xi}{2\pi} e^{i\sigma.\xi} \frac{-i}{2M\zf}\epsilon^{ij}_\perp \bigg[\dpi^j \bigg(F^\nu_{1,1} (x, \xi, t, \bfp^2, \bfd.\bfp) - 2 \zfs F^\nu_{1,3}(x, \xi, t, \bfp^2, \bfd.\bfp)\bigg)\bigg]\nonumber\\
&&- 2\zfs p^j_\perp F^{(e)\nu}_{1,2} (x, \xi, t, \bfp^2, \bfd.\bfp) + \frac{\xi }{M^2}\epsilon^{kl}_\perp \pp^k \dpi^l \dpi^j F^{(e)\nu}_{1,4} (x, \xi, t, \bfp^2, \bfd.\bfp) \bigg] \,,\\
\tilde{\rho}^{\nu j}_{UT}(x,\sigma,t,\bfp)&=&  \int^{\xi_{s}}_0 \frac{d \xi}{2\pi} e^{i\sigma.\xi}  \frac{-i}{M\zf}\epsilon^{ij}_\perp\bigg[ p^i_\perp H^{(e)\nu}_{1,1} (x, \xi, t, \bfp^2, \bfd.\bfp) +   \dpi^i H^\nu_{1,2}(x, \xi, t, \bfp^2, \bfd.\bfp)\bigg]\,, \label{rhoUT_H}
\ee
Note that, the first term of Eq.(\ref{RUTSiv_bt}) and in Eq.(\ref{RUTBM_bt}) have similar decomposition in terms of GTMDs for the Fourier transform in the $\bfb$ space.

For completeness, we list the analytical results of all the T-Odd leading twist GTMDs in LFQDM corresponding to the unpolarised and transversely polarised quark. 
The complete bilinear decompositions of the quark-quark correlator of Eqs.(\ref{WV_def},\ref{WT_def}) relate to the leading twist GTMDs reads as~\cite{Meissner:2009ww}
\be 
W^{\nu [\gamma^+]}_{[\lambda^{\prime\prime}\lambda^{\prime}]}&=&\frac{1}{2M} \bar{u}(P^{\prime\prime},\lambda^{\prime\prime})\bigg[ F_{1,1} + \frac{i\sigma^{i+} \pp^i}{P^+}F_{1,2} +  \frac{i\sigma^{i+} \dpi^i}{P^+}F_{1,3}+ \frac{i\sigma^{ij} \pp^i \dpi^j}{M^2}F_{1,4}\bigg] u(P^\prime,\lambda^\prime)\label{WV_def_App}\,,\\
W^{\nu [i \sigma^{j+}\gamma^5]}_{[\lambda^{\prime\prime}\lambda^{\prime}]}&=&\frac{1}{2M} \bar{u}(P^{\prime\prime},\lambda^{\prime\prime})\bigg[-\frac{i\epsilon^{ij}_\perp \pp^i }{M} H_{1,1} - \frac{i\epsilon^{ij} \dpi^i}{M}H_{1,2} + \frac{M i\sigma^{j+} \gamma^5}{P^+} H_{1,3} +  \frac{\pp^j i\sigma^{k+} \gamma^5 \bfp^k}{M P^+} H_{1,4} \nonumber \\
&& + \frac{\dpi^j i\sigma^{k+} \gamma^5 \bfp^k}{M P^+} H_{1,5}+ \frac{\dpi^j i\sigma^{k+} \gamma^5 \dpi^k}{M P^+} H_{1,6} +  \frac{\pp^j i\sigma^{+-} \gamma^5 }{M}H_{1,7}+  \frac{\dpi^j i \sigma^{+-} \gamma^5 }{M}H_{1,8}\bigg] u(P^\prime,\lambda^\prime)\,,\nonumber\\ \label{WT_def_App}
\ee
The model results for the T-Odd part of the leading twist GTMDs are as follows:
(i) for unpolarised quark with Dirac matrix structure $\Gamma=\gamma^+$:
\be 
F^{(o)\nu}_{1,1}(x,\xi,\bfd^2,\bfp^2,\bfd.\bfp) &=& N^\nu_{F11}\frac{1}{16\pi^3} \zf  \bigg[\left( \chi^{\prime}_1 - \chi^{\prime\prime}_1 \right) \Aodp \Aop + \bigg\{ \bfp^2 - \frac{\bfd^2}{4}\frac{(1-x)^2}{(1-\xi^2)}  \nonumber \\
&& + \frac{\xi (1-x)}{(1-\xi^2)}(\bfp.\bfd)\bigg\} \left( \chi^{\prime}_2 - \chi^{\prime\prime}_2 \right) \frac{\Atdp \Atp}{x^{\prime\prime} x^{\prime} M^2}\bigg] \exf\,, \label{F11} \\
F^{(o)\nu}_{1,3}(x,\xi,\bfd^2,\bfp^2,\bfd.\bfp) &=& N^\nu_{F13} \frac{1}{16\pi^3} \frac{(1-x)}{\zf}\frac{1}{2} \bigg[ \left( \chi^{\prime}_2 - \chi^{\prime\prime}_1 \right) \frac{\Aodp \Atp}{x^\prime(1+\xi)} + \left( \chi^{\prime}_1 - \chi^{\prime\prime}_2 \right) \frac{\Atdp \Aop}{x^{\prime\prime}(1-\xi)} \bigg] \nonumber \\
&&  \times \exf   +  \frac{1}{2(1-\xi^2)}  F^{(o)\nu}_{1,1}(x,\xi,\bfd^2,\bfp^2,\bfd.\bfp) \nonumber \\
&&+ \frac{1}{2M^2} \frac{\xi}{(1-\xi^2)}(\bfp.\bfd) F^{(o)\nu}_{1,4}(x,\xi,\bfd^2,\bfp^2,\bfd.\bfp)\,,\label{F13} \\
F^{(o)\nu}_{1,4}(x,\xi,\bfd^2,\bfp^2,\bfd.\bfp) &=& - N^\nu_{F14}\frac{1}{16\pi^3}  \frac{(1-x)}{\zf} \frac{1}{x^{\prime\prime}x^\prime}\left( \chi^{\prime}_2 - \chi^{\prime\prime}_2 \right)\Atdp\Atp \exf\,, \nonumber \\
\label{F14} 
\ee

(ii) for a transversely polarized quark with Dirac matrix structure $\Gamma=i\sigma^{j+} \gamma^5$: 
\be
H^{(o)\nu}_{1,2}(x,\xi,\bfd^2,\bfp^2,\bfd.\bfp) &=& N^\nu_{H12}\frac{1}{16\pi^3}  \zf \frac{1}{2}\bigg[ \left( \chi^{\prime}_2 - \chi^{\prime\prime}_1 \right) \frac{1-x^\prime}{x^\prime}\Aodp\Atp + \left( \chi^{\prime}_1 - \chi^{\prime\prime}_2 \right) \frac{1-x^{\prime \prime}}{x^{\prime \prime}}\Atdp\Aop \bigg] \nonumber \\
&&  \times \exf \,,\label{H12}
\ee
\be 
H^{(o)\nu}_{1,3}(x,\xi,\bfd^2,\bfp^2,\bfd.\bfp) &=&N^\nu_{H13}\frac{1}{16\pi^3} \frac{1}{\zf} \bigg[ \left( \chi^{\prime}_1 - \chi^{\prime\prime}_1 \right) \Aodp \Aop \nonumber\\
&& + \bigg(\bfp^2 - \frac{(1-x)^2}{(1-\xi^2)} \frac{\bfd^2}{4} \bigg) \left( \chi^{\prime}_2 - \chi^{\prime\prime}_2 \right) \frac{\Atdp \Atp}{x^\prime x^{\prime\prime} M^2}  \bigg] \exf  \nonumber \\
&& -  \frac{1}{2M^2(1-\xi^2)}  \bfd^2 H^{(o)\nu}_{1,2}(x,\xi,\bfd^2,\bfp^2,\bfd.\bfp) \,,\label{G13}\\
H^{(o)\nu}_{1,4}(x,\xi,\bfd^2,\bfp^2,\bfd.\bfp) &=& -N^\nu_{H14}\frac{1}{16\pi^3}  \frac{1}{ \zf} \left( \chi^{\prime}_2 - \chi^{\prime\prime}_2 \right) \frac{2}{x^{\prime\prime}x^\prime} \Atdp \Atp  \exf\,,\nonumber \\ \label{H14}\\
H^{(o)\nu}_{1,5}(x,\xi,\bfd^2,\bfp^2,\bfd.\bfp) &=& N^\nu_{H15} \frac{1}{16\pi^3} \bigg[ \frac{\xi}{ \zfs^{3/2}} \frac{(1-x)}{x^{\prime\prime}x^\prime} \left( \chi^{\prime}_2 - \chi^{\prime\prime}_2 \right) \Atdp \Atp \bigg] \nonumber \\
&& \times  \exf \nonumber \\
&& + \frac{\xi}{(1-\xi^2)} H^{(o)\nu}_{1,7}(x,\xi,\bfd^2,\bfp^2,\bfd.\bfp) \,, \label{H15}\\
H^{(o)\nu}_{1,6}(x,\xi,\bfd^2,\bfp^2,\bfd.\bfp) &=& N^\nu_{H16} \frac{1}{16\pi^3} \bigg[ \frac{1}{ \zfs^{3/2}} \frac{(1-x)^2}{2 x^{\prime\prime}x^\prime} \left( \chi^{\prime}_2 - \chi^{\prime\prime}_2 \right) \Atdp \Atp \bigg] \nonumber \\
&& \times   \exf  \nonumber \\
&& + \frac{1}{2(1-\xi^2)} H^{(o)\nu}_{1,2}(x,\xi,\bfd^2,\bfp^2,\bfd.\bfp) 
+ \frac{\xi}{(1-\xi^2)}  H^{(o)\nu}_{1,8}(x,\xi,\bfd^2,\bfp^2,\bfd.\bfp)\,, \label{H16}\\
H^{(o)\nu}_{1,7}(x,\xi,\bfd^2,\bfp^2,\bfd.\bfp) &=&  - N^\nu_{H17}\frac{1}{16\pi^3}  \zf  \frac{1}{2}\bigg[ \left( \chi^{\prime}_2 - \chi^{\prime\prime}_1 \right) \frac{1}{x^\prime}\Aodp\Atp - \left( \chi^{\prime}_1 - \chi^{\prime\prime}_2 \right) \frac{1}{x^{\prime \prime}}\Atdp\Aop \bigg] \nonumber \\
&&  \times \exf\,,\label{H17}\\
H^{(o)\nu}_{1,8}(x,\xi,\bfd^2,\bfp^2,\bfd.\bfp) &=& N^\nu_{H18}\frac{1}{16\pi^3}  \zf \frac{1}{4}\bigg[ \frac{(1-x^\prime)}{x^\prime} \left( \chi^{\prime}_2 - \chi^{\prime\prime}_1 \right) \Aodp\Atp \nonumber \\
&& - \left( \chi^{\prime}_1 - \chi^{\prime\prime}_2 \right) \frac{(1-x^{\prime \prime})}{x^{\prime \prime}}\Atdp\Aop \bigg]   \exf \,,\label{H18}
\ee
The normalization constants $N^\nu_{\Lambda\lambda}$ are 
\be 
N^\nu_{F11},N^\nu_{G11} N^\nu_{H11}, N^\nu_{H12} 
&=&  \bigg(C^2_SN^2_S+C^2_A \big(\frac{1}{3}N^2_0+\frac{2}{3}N^2_1\big)\bigg)^\nu\,, \nonumber\\
N^\nu_{F14}, N^\nu_{G14} ,  N^\nu_{H17}, N^\nu_{H18}
&=&\bigg(C^2_SN^2_S+C^2_A\big(\frac{1}{3}N^2_0-\frac{2}{3}N^2_1\big)\bigg)^\nu\,, \nonumber\\
N^\nu_{F12}, N^\nu_{F13}, N^\nu_{G12}, N^\nu_{G13} , N^\nu_{H13}, N^\nu_{H14}\,, N^\nu_{H15},N^\nu_{H16} 
&=&  \bigg(C^2_SN^2_S-C^2_A \frac{1}{3}N^2_0\bigg)^\nu\,,
\ee

{\bf Data Availability Statement} No Data associated in the manuscript.

\bibliography{bib_WD_ZetaOdd_fR1}
\end{document}